\documentclass{article}
\PassOptionsToPackage{numbers, compress}{natbib}

\usepackage[preprint]{neurips_2025}

\usepackage[utf8]{inputenc} 
\usepackage[T1]{fontenc}    
\usepackage{url}            
\usepackage{booktabs}       
\usepackage{amsfonts}       
\usepackage{nicefrac}       
\usepackage{microtype}      
\usepackage{xcolor}         
\usepackage{enumitem}
\usepackage{soul}

\author{
Nina S. Nellen,\textsuperscript{1,$\dagger$, *}
Polina Turishcheva,\textsuperscript{1,$\dagger$,*}
Michaela Vystrčilová,\textsuperscript{1}
Shashwat Sridhar,\textsuperscript{2,3}
\\ \textbf{Tim Gollisch,\textsuperscript{2--5}
Andreas S. Tolias,\textsuperscript{6--9}
Alexander S. Ecker\textsuperscript{1,10,*}}
\\\\ \textsuperscript{1} \footnotesize Institute of Computer Science and Campus Institute Data Science, University G{\"o}ttingen, Germany
\\ \textsuperscript{2} \footnotesize University Medical Center G{\"o}ttingen, Department of Ophthalmology, G{\"o}ttingen, Germany
\\ \textsuperscript{3} \footnotesize Bernstein Center for Computational Neuroscience G{\"o}ttingen, G{\"o}ttingen, Germany
\\ \textsuperscript{4} \footnotesize Cluster of Excellence ``Multiscale
Bioimaging: from Molecular Machines to Networks of Excitable Cells'' 
\\ \footnotesize (MBExC), University of G{\"o}ttingen, G{\"o}ttingen, Germany
\\ \textsuperscript{5} \footnotesize Else Kr{\"o}ner Fresenius Center for Optogenetic Therapies,  University Medical Center G{\"o}ttingen, G{\"o}ttingen, Germany
\\\textsuperscript{6} \footnotesize Department of Ophthalmology, Byers Eye Institute, Stanford University School of Medicine, Stanford, CA, US
\\\textsuperscript{7} \footnotesize Stanford Bio-X, Stanford University, Stanford, CA, US
\\\textsuperscript{8} \footnotesize Wu Tsai Neurosciences Institute, Stanford University, Stanford, CA, US
\\\textsuperscript{9} \footnotesize Department of Electrical Engineering, Stanford University, Stanford, CA, US
\\ \textsuperscript{10} \footnotesize Max Planck Institute for Dynamics and Self-Organization, G{\"o}ttingen, Germany
\\\\
\textsuperscript{$\dagger$} \footnotesize Shared contribution
\\
\textsuperscript{$\star$}\texttt{ninasophie.nellen@stud.uni-goettingen.de}
\\\textsuperscript{$*$}\texttt{\{turishcheva,ecker\}@cs.uni-goettingen.de}
}

\usepackage{graphicx} 
\usepackage{algorithm}
\usepackage{algpseudocode}

\usepackage[english]{babel}
\usepackage[hidelinks]{hyperref}

\usepackage{amsmath,amssymb,amsthm, mathtools}
\usepackage{enumerate}  
\usepackage{mathdots}
\usepackage{yhmath}
\usepackage{pythontex} 
\usepackage{epsfig}
\usepackage{booktabs}
\usepackage{dsfont}
\usepackage{cleveref}
\crefname{algorithm}{Alg.}{Algs.}
\crefname{figure}{Fig.}{Figs.}
\crefname{equation}{Eq.}{Eqs.}
\crefname{appendix}{App.}{Apps.}
\crefname{section}{Sec.}{Secs.}

\usepackage{tikz}
\usepackage{graphicx}
\usepackage{subcaption}
\usepackage{wrapfig}
\usepackage{csquotes}

\numberwithin{equation}{section}

\title{Learning to cluster neuronal function}

\begin{document}

\maketitle

\begin{abstract}

Deep neural networks trained to predict neural activity from visual input and behaviour have shown great potential to serve as digital twins of the visual cortex.
Per-neuron embeddings derived from these models could potentially be used to map the functional landscape or identify cell types. 
However, state-of-the-art predictive models of mouse V1 do not generate functional embeddings that exhibit clear clustering patterns which would correspond to cell types.
This raises the question whether the lack of clustered structure is due to limitations of current models or a true feature of the functional organization of mouse V1. 
In this work, we introduce DECEMber -- Deep Embedding Clustering via Expectation Maximization-based refinement -- an explicit inductive bias into predictive models that enhances clustering by adding an auxiliary $t$-distribution-inspired loss function that enforces structured organization among per-neuron embeddings.
We jointly optimize both neuronal feature embeddings and clustering parameters, updating cluster centers and scale matrices using the EM-algorithm.
We demonstrate that these modifications improve cluster consistency while preserving high predictive performance and surpassing standard clustering methods in terms of stability.
Moreover, DECEMber generalizes well across species (mice, primates) and visual areas (retina, V1, V4). 
The code is available at \url{https://github.com/Nisone2000/sensorium/tree/neuroips_version}.
\end{abstract}

\section{Introduction}

Understanding whether neurons form discrete cell types or lie on a continuum is a fundamental question in neuroscience \cite{zeng2022cell}. Previous research has extensively investigated the morphological and electrophysiological properties of neurons in the visual cortex.
While discrete anatomical and transcriptomic classifications have been proposed ~\cite{defelipe2013new,oberlaender2012cell,markram2015reconstruction}, recent work on the mouse brain suggests a more continuous organization \cite{scala2019layer, Weis2025unsupervised}.
Significantly less attention has been devoted to the neurons' functional properties.
Each neuron can be characterized by a function that maps high-dimensional sensory inputs to its one-dimensional neuronal response. 
These functions are highly non-linear, making their analysis complex. 
Discrete functional cell types are well established in the retina~\cite{baden2016functional} but their existence remains unclear in the mouse visual cortex.

Recently, deep networks showed great potential for predicting neural activity from sensory input \cite{sensorium, turishcheva2024dynamic, Antolik, schmidt2025modelingdynamicneuralactivity, Wang2023.03.21.533548, turishcheva202, franke2021behavioral, ecker2018rotation, Vystrcilova2024} and also in inferring novel functional properties \cite{walker2019inception, franke2022state, hofling2024chromatic}.
These networks learn per-neuron vectors of parameters, which are interpreted as neuronal functional embeddings.
There were several attempts to use these embeddings to reveal the underlying structure of neuronal population functions through unsupervised clustering \cite{Tong2023.11.03.565500, ivan, turishcheva202, burg2024discriminativestimulifunctionalcell}. 
However, in none of these studies well-separated clusters emerged, raising the question of whether distinct functional cell types exist among excitatory neurons in the mouse visual cortex. 
A central challenge is cluster consistency: How reliably are neurons grouped into the same cluster across different model runs? 
Clustering metrics such as the Adjusted Rand Index (ARI) \cite{ARI}, which evaluates cluster assignment agreement across different seeds or clustering methods
and similarity metrics between individual neurons’ remained relatively low \cite{turishcheva202}.
These low scores show that clustering results lack the stability and distinctiveness necessary to making strong claims about biological interpretations.

In this work, we incorporate an explicit clustering bias into the training of neuronal embeddings to improve the identifiability of functional cell types,
One could view it as model-driven hypothesis testing: if clear functional cell types exist then such bias should improve the model performance, embeddings structure and/or cluster consistencies.

To improve the cluster separability of neuronal embeddings we took inspiration from Deep Embedding Clustering (DEC) \cite{DEC} and introduced a new clustering loss, which combines updating clusters' locations and shapes along with learning feature representations. 
We measured the consistency of clustered features across models fitted on different seeds by computing ARI on their clustering results.
Additionally, we examined how the clustering loss strength influenced models' performance.

Our contributions are

\begin{itemize}[noitemsep, topsep=-5pt, leftmargin=*, itemsep=0pt, partopsep=5pt]
    \item We adapted the DEC-loss  \cite{DEC}  to allow for non-isotropic multivariate clusters of different sizes by learning a multivariate $t$ mixture model \cite{tMM}.
    \item We improved cluster consistency while maintaining a state-of-the-art predictive model performance.
    \item We showed that our method generalizes well, improving cluster consistency across different species,  visual areas, and model architectures.
\end{itemize}

\section{Background and related work}

\paragraph{Predictive models for visual cortex.} 
In comparison to task-driven networks \cite{yamins2014performance, cadieu2014deep, cadena2019deep, pogoncheff2023explaining}, pioneering data-driven population models \cite{Antolik,Batty2016MultilayerRN,mcintosh2016deep} introduced the core-readout framework, which separates the stimulus-response functions of neurons into a shared nonlinear feature space (core) and per-neuron specific set of linear weights -- the readout. The core is shared among all neurons and outputs a nonlinear set of basis functions spanning the feature space of the neuronal nonlinear input-output functions of dimension (height $\times$ width $\times$ feature channels).
The early models were extended by including behavioral modulation \cite{sinz2018stimulus, franke2022state}, latent brain state \cite{schmidt2025modelingdynamicneuralactivity, NEURIPS2021_84a529a9} or the perspective transformations of the eye \cite{Wang2023.03.21.533548}.
The core architecture was improved by introducing biological biases such as a rotation-equivariant core \cite{ecker2018rotation} to account for orientation selectivity in V1 neurons \cite{Tan2011}, extending to dynamic models with video input \cite{sinz2018stimulus, turishcheva2024dynamic, Wang2023.03.21.533548, hofling2024chromatic, Vystrcilova2024} or using transformer architectures \cite{li2023v1t}. 

Klindt et al.~\cite{whatandwhere} introduced a factorized readout for each neuron, comprising a spatial mask \( M_n \) specifying its receptive field (RF) position and feature weights. This approach was refined by Lurz et al.~\cite{Lurz}, who proposed the Gaussian readout, replacing the full spatial mask with a pair of coordinates \((x_n, y_n)\) drawn from a learned normal distribution. 
To predict the neuronal response the model computes the dot product between the neuron's weight vector (per-neuron embedding) and each feature map at the RF location. 
For later visual layers, like V4, the receptive field location is not necessarily fixed. 
Therefore, Pierzchlewicz et al. \cite{pierzchlewicz2023energy} introduced an attention readout, which indicates the most important feature locations for a neuron $n$ depending on the input image.

While different readouts exist, few works have examined their consistency. 
Turishcheva et al.~\cite{turishcheva202} showed that factorized readouts produced more consistent neuronal clusters than Gaussian readouts, despite lower predictive performance. 
They addressed this by introducing adaptive log-norm regularization to balance model expressiveness and feature consistency.
However, the ARI scores were still not high enough to claim distinct cell types. Moreover, their work involved a rotation-equivariant convolutional core and required a post-hoc alignment procedure~\cite{ustyuzhaninov2019rotation} to interpret the cluster structures.



\paragraph{Deep embedding clustering.} 
Deep Embedding Clustering (DEC) \cite{DEC} combines clustering with representation learning. 
It introduced a clustering loss that simultaneously drives learning the cluster centroids and encourages the feature representation to separate the clusters.
After pretraining a deep autoencoder without the clustering loss, the cluster centers $\mu_j$ are initialized using k-means \cite{kmeans}. 
DEC then minimizes a Kullback-Leibler (KL) divergence of soft cluster assignments $Q$ and an auxiliary target distribution $P$ defined as follows:

\begin{minipage}{0.48\textwidth}
\vspace{-12pt} 
\begin{equation}
\vspace{-12pt}
q_{ij} = \frac{\left(1 + \frac{\lVert z_i - \mu_j \rVert^2}{\nu} \right)^{-\frac{\nu + 1}{2}}}
{\sum_{j'} \left(1 + \frac{\lVert z_i - \mu_{j'} \rVert^2}{\nu} \right)^{-\frac{\nu + 1}{2}}}
\label{eq:qij}
\end{equation}
\end{minipage}
\hfill
\begin{minipage}{0.48\textwidth}
\begin{equation}
p_{ij} = \frac{q_{ij}^2 / f_j}{\sum_{j'} q_{ij'}^2 / f_{j'}} \quad \text{with } f_j = \sum_i q_{ij}
\label{eq:pij}
.\end{equation}
\end{minipage}
\vspace{0.2cm}

The $q_{ij}$s are the probabilities of sample $z_i$ belonging to cluster $j$ and are represented by a Student's $t$-distribution with unit scale and degree of freedom $\nu$ being set to 1. The target distribution $P$ (\Cref{eq:pij}) is chosen such that it:
\begin{itemize}[noitemsep, leftmargin=*, itemsep=0pt]
    \item \textbf{``strengthens predictions.''} 
    Original values $q_{ij}$ denote the soft assignment probability of a data point $i$ belonging to cluster $j$.
    Squaring $q_{ij}$ and then re-normalizing makes high-confidence assignments more dominant while further diminishing the influence of low-confidence ones.
    \item \textbf{``emphasizes high-confidence data points.''} A high $q_{ij}$ dominates $q_{ij}^2/f_j$, meaning that points strongly associated with a cluster contribute more to $p_{ij}$. 
    \item \textbf{``normalizes loss contribution of each centroid to prevent large clusters from distorting the hidden feature space.''} Without $f_j$, larger clusters could dominate the feature space since they would contribute disproportionately to the loss. By dividing by $f_j$ the impact of each cluster is normalized, ensuring that smaller clusters are not overshadowed by larger ones.
\end{itemize}


\begin{figure}[t]
    \centering
    \includegraphics[width=\textwidth]{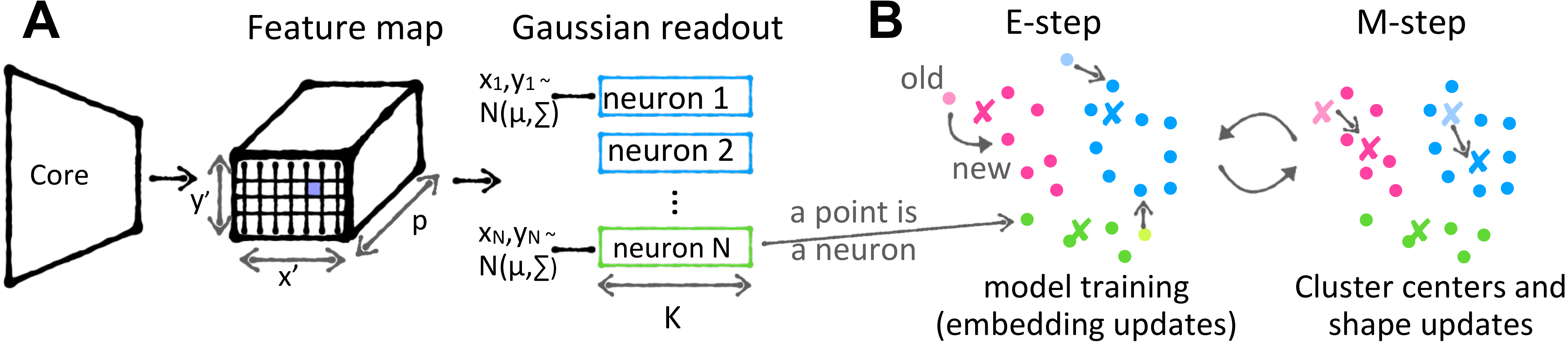}
    \caption{\textbf{A: Model architecture:} The model consists of a neuronwise shared core outputting a feature map of size (height $\times$ width $\times$ feature channels) and neuron specific Gaussian readouts. They consist of a receptive field position and a weight vector. The RF position chooses the vector in the feature map which is then combined with the neuron's weight vector by a dot product to get the neuron's response. \textbf{B: Clustering procedure: } We're clustering the readouts with an additional loss to incorporate the cluster bias into the features. We update the clustering parameters (cluster centers and scale matrices) with an EM step of a $t$ mixture model as in \cref{alg:model_training}. }
    \label{fig:model}
\end{figure}

\section{DECEMber -- Deep Embedding Clustering via Expectation Maximization-based refinement} 
 
DECEMber jointly trains a neural response predictor and learns a clustering structure by optimizing a Deep Embedding Clustering–inspired loss, with cluster parameters updated via the EM algorithm.
We now describe our approach (illustrated in \Cref{fig:model}, described in \cref{alg:model_training}).

\paragraph{Predictive model for visual cortex.} 
We build on a state-of-the-art predictive model \cite{sensorium} for responses $r_i$ of neurons $i=1,...,N$ to visual stimuli $s \in \mathbb{R}^{H' \times W' \times T \times C}$. 
Here $H'$ and $W'$ are height and width of the input, $T$ time if the input is a video and $C$ is the amount of channels:  $C=1$ for grayscale or $C=3$ for RGB.
For static visual input (images), $T=1$ and could be ignored.
If behavior variables  -- such as pupil size, locomotion 
speed, and changes in pupil size -- are present, they are concatenated to the stimuli as channels \cite{sensorium}. 
The model combines a shared convolutional core $\Phi$ with neuron-specific Gaussian readouts $\psi_i$ (\Cref{fig:model}A). 
The core outputs a feature space $\Phi(s) \in \mathbb{R}^{H \times W \times K}$. 
We denote the core's parameters by $\theta$. 
The readout \cite{Lurz} $\psi_i: \mathbb{R}^{H \times W \times K} \mapsto \mathbb{R}$ first selects the features from $\Phi$ at the neuron's receptive field location $(x_i, y_i)$ using bilinear interpolation, which we write with a slight abuse of notation as $\Phi(\tilde s)_{x_iy_i} \in \mathbb{R}^K$, resulting in a feature vector $\phi_i \in \mathbb{R}^K$. 
It then computes the predicted neuronal response $\hat r_i = z_i^T \phi_i$, where $z_i \in \mathbb{R}^K$ is the neuron-specific readout weight (its functional embedding), overall
\begin{equation}
    \hat r_i( s) = \psi_i(\Phi( s)) = z_i^T \Phi(\tilde s)_{x_iy_i}.
\end{equation}

\paragraph{Clustering loss on readout weights.}
We encourage a well-clustered structure on the neuron-specific readout weights $Z = [z_1, \dots, z_N]$ by incorporating a clustering objective directly into training. Specifically, we augment the standard model loss with a clustering loss that minimizes the KL divergence between soft cluster assignments $q_{ij}$, (\cref{eq:soft_assignments}) and targets $p_{ij}$, (\cref{eq:pij}), 
\begin{equation}
    L_{\mathrm{cluster}} = KL(Q(Z) || P(Z)) = \sum_{i=1}^N \sum_{j=1}^J p_{ij} \log{\left( \frac{p_{ij}}{q_{ij}}\right)}.
    \label{eq:KL}
\end{equation}
This auxiliary loss encourages the embeddings to form $J$ distinct clusters.

\begin{algorithm}[t]
    \caption{Model Training with clustering loss} \label{alg:model_training}
    \begin{algorithmic}
        \State \textbf{Inputs:} Degrees of freedom $\nu$, clustering weight $\beta$, core parameters $\theta$, neuronal embeddings (readout) $Z$
        \State \textbf{Output:} Parameters $ \mu_j, \Sigma_j$, $\theta$ and $Z$

        \State \textbf{Pretraining:} Train the predictive model by optimizing $L_{\text{model}}$ w.r.t. $\theta$ and $Z$ for $m$ epochs
        \State \textbf{Initialize:} Cluster centers $\mu_j$ with $k$-means and diagonal scale matrix $\Sigma_j$ as within-cluster variance
        
        \For{epoch $t = 1$ to $T$}
            \For{minibatch $b$ in dataset}
                \State \textbf{(1) E-step (Expectation):} Compute
                \State \quad 1.1 Soft assignments $q_{ij} = \frac{ f_{t}(z_i ; \mu_j, \Sigma_j, \nu )}
{\sum_{j'=1}^J f_{t}(z_i ; \mu_{j'}, \Sigma_{j'}, \nu )}$  
                \hfill \refstepcounter{equation}(\theequation)\label{eq:soft_assignments}
                \State \quad 1.2 Latent scales 
                $ u_{ij} = \frac{\nu + K}{\nu + (z_i - \mu_j)' \Sigma_j^{-1} (z_i - \mu_j) } $
                \hfill \refstepcounter{equation}(\theequation)\label{eq:uij}\vspace{3pt}

                \State \textbf{(2) M-step (Maximization):} Update parameters
                \State \quad 2.1 Update $ \mu_j = \frac{\sum_{i=1}^N q_{ij} u_{ij} \, z_i }{\sum_{i=1}^N q_{ij} u_{ij} } $
                \hfill \refstepcounter{equation}(\theequation)\label{eq:muj}
                \State \quad 2.2 Update  $ \Sigma_j = \frac{\sum_{i=1}^N q_{ij} u_{ij} (z_i - \mu_j)(z_i - \mu_j)'} {\sum_{i=1}^N q_{ij}} $ 
                \hfill \refstepcounter{equation}(\theequation)\label{eq:sgimaj}\vspace{3pt} 

                \State \textbf{(3) Gradient step: }Optimize predictive model parameters
                \State \quad 3.1 Minimize $L= L_{\text{model}} + \beta KL(Q||P) $ w.r.t $\theta, Z$
                \State \quad with $p_{ij} = \frac{q_{ij}^2 / f_j}{\sum_{k} q_{ik} / f_{k}}$ and $f_j = \sum_i q_{ij}$\\
            \EndFor
        \EndFor \\
        \vspace{3pt}
        \Return $\mu, \Sigma, \theta, Z$
    \end{algorithmic}
\end{algorithm}

Xie et al. \cite{DEC} use a pretrained autoencoder with well-separated embeddings and model soft cluster assignments using a Student's $t$-distribution with the learnt cluster centers and fixed unit scale. 
This setup is too constrained for our regression setting where the mean and the scale of the embeddings $z_i$ is not a free parameter but restricted by the regression loss. 
We therefore adopt a more flexible approach by using a multivariate Student's $t$-mixture model (TMM) \cite{tMM}, where each cluster is not only defined by its center $\mu_j$ but also adapts its scale matrix $\Sigma_j$. 
We update these parameters during training using the Expectation-Maximization (EM) algorithm (described next).


\paragraph{EM step to update cluster parameters.} 

Instead of directly learning the cluster centroids via gradient descent, we updated them after each batch using the EM algorithm applied to the Student’s $t$-mixture model, 
$f_{\text{TMM}}(z_i; \Theta) = \frac{1}{J} \sum_{j=1}^J f_t(z_i; \mu_j, \Sigma_j, \nu),
$ \cite{tMM} where degree of freedom $\nu$ controls the probability mass in the tails (if $\nu \to \infty$ the $t$-distribution becomes Gaussian). 
The density of the multivariate Student's $t$-distribution is:
\begin{align}
f_t(z_i ; \mu_j, \Sigma_j, \nu) &= \frac{\Gamma\left(\frac{\nu + K}{2}\right)}{
\Gamma\left(\frac{\nu}{2}\right) 
\nu^{\frac{p}{2}} 
\pi^{\frac{p}{2}} 
|\Sigma_j|^{\frac{1}{2}}} 
\left(1 + \frac{1}{\nu} 
(z_i - \mu_j)^T 
\Sigma_j^{-1} 
(z_i - \mu_j)
\right)^{-\frac{\nu + K}{2}} \label{eq:student-t-closed} \\
&= \int_0^\infty 
\mathcal{N}(z_i \mid \mu_j, \tfrac{1}{u} \Sigma_j) 
\cdot 
\mathrm{Gamma}\!\left(u \,\middle|\, \tfrac{\nu}{2}, \tfrac{\nu}{2} \right)
\, du.\, \label{eq:student-t-integral} 
\end{align}
with the latter being the so-called shape-rate form of the $t$-distribution \cite{mclachlan2000finite}.
This interpretation is useful because introducing the Gamma-distributed latent variable $u$ allows closed-form M-step updates for $\mu_j$ and $\Sigma_j$, whereas direct likelihood optimization in a $t$-mixture model does not generally admit closed-form solutions. 

The full procedure is summarized in \Cref{alg:model_training} and alternates between: (1)~E-Step: Compute soft cluster assignments $q_{ij}$ (\cref{eq:soft_assignments}) -- the probability of feature $i$ belonging to cluster $j$ -- and the latent scaling factors $u_{ij}$ (\cref{eq:uij}).
(2)~M-Step: Update cluster means $\mu_j$ (\cref{eq:muj}) and (diagonal) scale matrices $\Sigma_j$ (\cref{eq:sgimaj}). (3)~Gradiet step: Update the parameters of the core and readout via one iteration of stochastic gradient descent.

\section{Experiments}



\paragraph{Clustering loss hyperparameters.}
\label{sec:beta}
For the clustering loss, we fixed the degrees of freedom to $\nu = 2.1$, just above the threshold where the variance $\frac{\nu}{\nu - 2} \Sigma$ becomes defined (only for $\nu > 2$). 
To balance model flexibility and robustness, we allowed each cluster to have its own diagonal scale matrix $\Sigma_j$, which alloed for different variances per embedding dimension while preventing overfitting.
For each dataset, we adjusted the clustering strength $\beta \in \mathbb{R}$ such that it is in the same order of magnitude as the model loss at initialization. 

\paragraph{Pretraining and cluster initialization.}
Before adding the clustering loss, we pretrained the baseline model for $m$ epochs, such that the model could already predict the responses reasonably well. 
We explored $m=5, \dots, 40$ to assess how the length of pretraining (PE) affected our results.  
We followed Turishcheva et al. \cite{turishcheva202}
for the pretraining procedure,  minimizing the following loss:
\begin{equation}
    L_{\mathrm{model}} = L_{\mathrm{P}} + L_{\mathrm{reg}} = \frac{1}{N} \sum_{l=1}^L\sum_{i=1}^{N} \left(\hat{r}_{il} -r_{il}  \log \hat{r}_{il} \right)+ L_{\mathrm{reg}}
    \label{L_model}
\end{equation}
where $L_{\mathrm{P}}$ is the Poisson loss that aligned per-image $l= 1, \dots, L$ model predictions $\hat r_{il}$ with observed neuronal responses $r_{il}$ since neuron's firing rates follow a Poisson process \cite{Poisson}, and $L_{\mathrm{reg}}$ is the adaptive regularizer that was shown to result in improved embedding consistency \cite{turishcheva202}.

After pretraining, we initialized the cluster centroids $\mu_j$ with k-means \cite{kmeans} and the diagonal scale matrices $\Sigma_j$ as the within-cluster variances. 
We continued training using $L_{model} + \beta L_{\mathrm{cluster}}$, with $L_{cluster}$ as in \cref{eq:KL} and 
scaled with $\beta$.

\paragraph{Evaluation of model performance.} 
Building on previous research \cite{corr1, ecker2018rotation, walker2019inception, burg2021learning, sensorium, li2023v1t}, we evaluated the model's predictive performance by computing the Pearson correlation (across images in the test set) between the measured and predicted neural responses, averaged across neurons. 

\paragraph{Evaluation of embedding consistency.} 
We wanted to assess the relative structure of the embedding space: Do the same groups of neurons consistently cluster together across models fit with different initial conditions?
To quantify this notion, we took DECEMber's cluster assignments and measured how often neuron pairs are assigned to the same group using the Adjusted Rand Index (ARI) \cite{ARI}, which quantifies the similarity between two clustering assignments, $X$ and $Y$.
The ARI remains unchanged under permutations of cluster labels. 
ARI equals one if and only if the two partitions are identical and it equals zero when the partitions agreement is no better than random. 

To compare DECEMber with a baseline, we extracted neuronal embeddings from the fully converged default model and fitted Gaussian Mixture Models (GMMs) using the same number of clusters as DECEMber, diagonal covariance, and a regularization of $10^{-6}$.
We then computed the ARI across three GMM partitions from baseline models initialized with different seeds, using a fixed GMM seed.

\paragraph{Visualization.} To visualize the neuronal embeddings, we employed t-SNE \cite{tSNE}, following the guidelines of \cite{tsne_setting}. 
Specifically, we set the perplexity to $N/100$, the learning rate to 1 and early exaggeration to $N/10$. 
To be comparable with prior work \cite{ivan,turishcheva202}, we randomly sample 2,000 neurons from each of the seven mice in the dataset and used the same neurons across all visualizations.

\section{Results}

\paragraph{DEC-loss needs learned scale: toy example illustration.} 
To assess whether the DEC-loss provides a useful clustering when applied to model weights instead of autoencoder embeddings, we constructed a simple toy example consisting of linear neurons whose responses are given as $y_i = z_i^T x$, where $z_i$ are the neurons' weights and $x$ the stimulus. 
We generated 8000 stimuli, each of the form $x = (x_1, \dots x_K) \in \mathbb{R}^{30}$ with $x_k \sim \mathcal{U}(-1,1)$.
For each stimulus $x$, we created responses $y = (y_1, \dots y_N) \in \mathbb{R}^{N}$ of $N=2500$ neurons. 
The neurons were split into two groups: for 2000 of them, $z_i = 1 + \epsilon_{i1}$ with $\epsilon_{i1} \sim \mathcal{U}(-1/300,1/300)$; for the other 500, $z_i = 1.5 + \epsilon_{i2}$ with $\epsilon_{i2} \sim \mathcal{U}(-1/120,1/120)$. 
We pretrained a linear regression model on this data 
for 25 epochs by minimizing the MSE. 
After that we continued training, minimizing only the KL divergence on the learned centers using the DEC-loss versus DECEMber. 
In theory the model should learn the weights of the two clusters centered at $\mu_1 = (1, \dots,1) \in \mathbb{R}^{30}$and $\mu_2 = (1.5,\dots, 1.5) \in \mathbb{R}^{30}$ and assign 2000 neurons to cluster 1 and 500 neurons to cluster 2.

\begin{figure}[t!]
    \begin{subfigure}[t]{0.28\textwidth}
        \begin{tikzpicture}
            \node[anchor=north west, inner sep=0] (img) at (0,0)
                {\includegraphics[width=\textwidth]{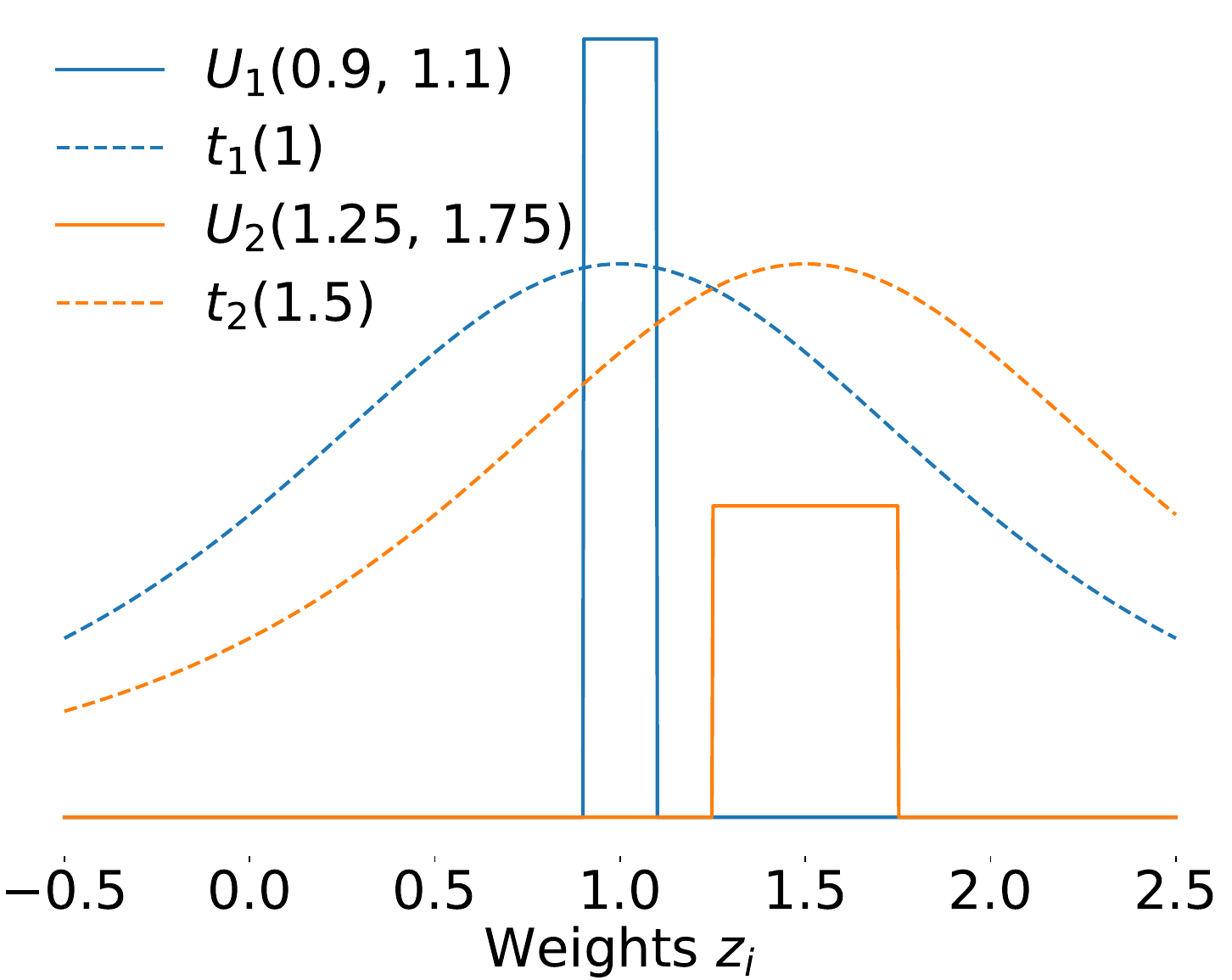}};
            \node[anchor=north west, inner sep=0pt, xshift=0em, yshift=0.3em] at (img.north west)
                {{\fontfamily{qhv}\selectfont\bfseries A}};
        \end{tikzpicture}
    \end{subfigure}%
    \hspace{-5mm}
    \begin{subfigure}[t]{0.22\textwidth}
        \begin{tikzpicture}
            \node[anchor=north west, inner sep=0pt] (img) at (-50,0)
                {\includegraphics[width=\textwidth]{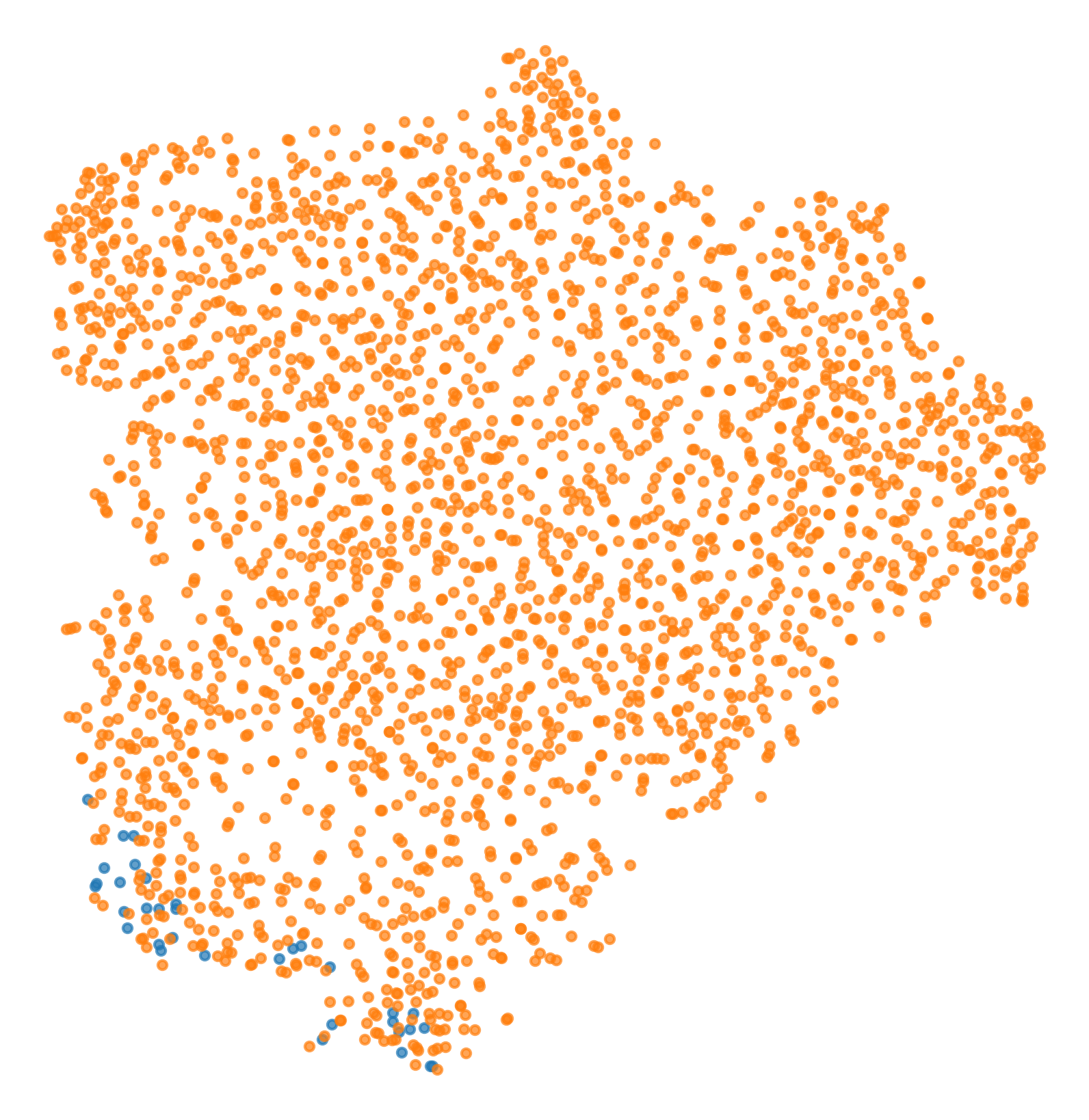}};
            \node[anchor=north west, inner sep=1pt, xshift=-0.5em, yshift=0.4em] at (img.north west)
                {{\fontfamily{qhv}\selectfont\bfseries B}};
        \end{tikzpicture}
    \end{subfigure}%
    \begin{subfigure}[t]{0.22\textwidth}
        \begin{tikzpicture}
            \node[anchor=north west, inner sep=0pt] (img) at (-2,0)
                {\includegraphics[width=\textwidth]{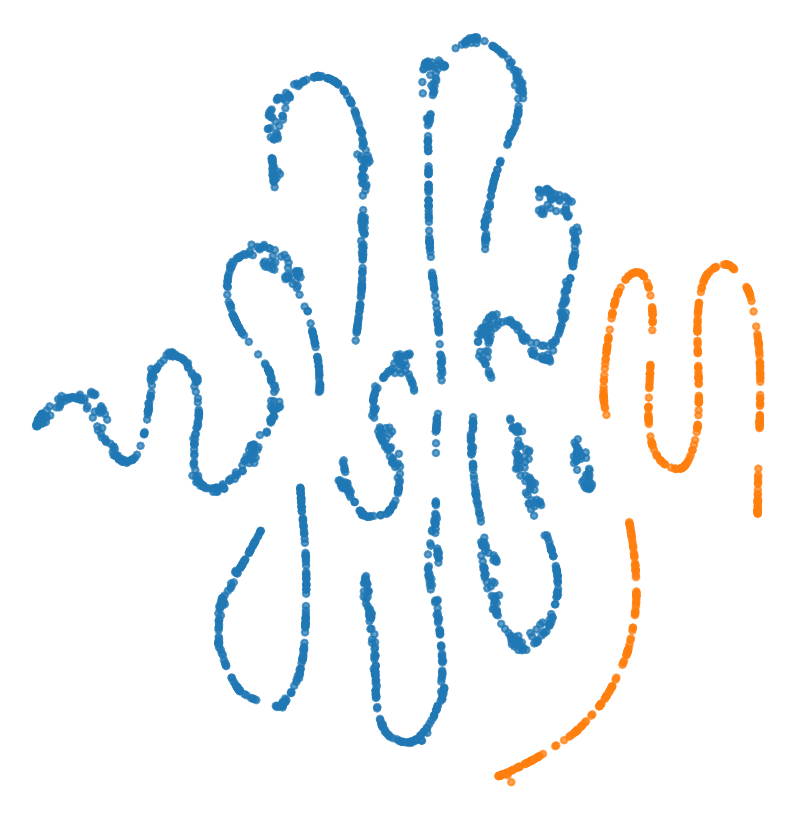}};
            \node[anchor=north west, inner sep=1pt, xshift=-0.5em, yshift=0.4em] at (img.north west)
                {{\fontfamily{qhv}\selectfont\bfseries C}};
        \end{tikzpicture}
    \end{subfigure}%
    \hspace{2mm}
    \begin{subfigure}[t]{0.28\textwidth}
        \begin{tikzpicture}
            \node[anchor=north west, inner sep=0] (img) at (0,0)
                {\includegraphics[width=\textwidth]{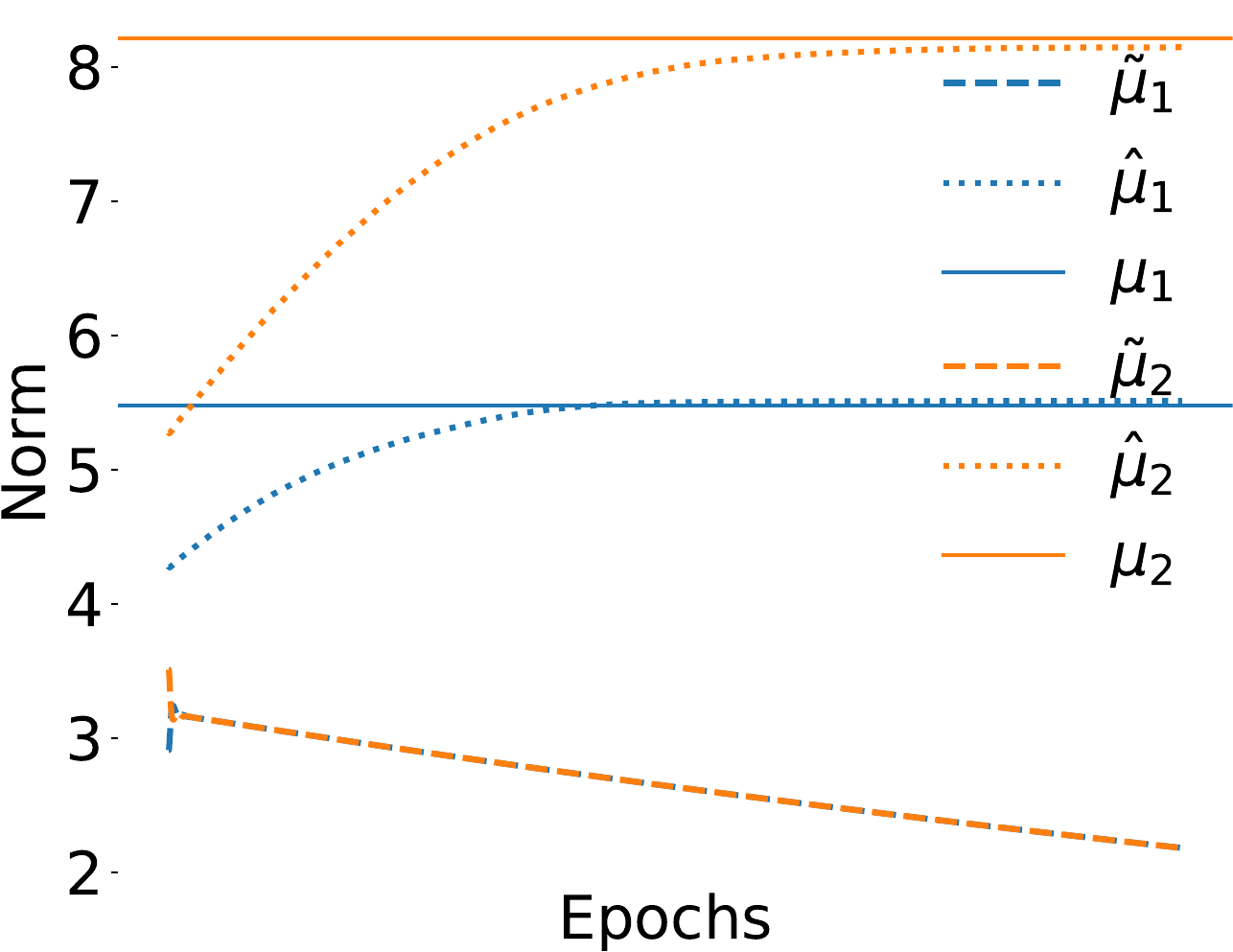}};
            \node[anchor=north west, inner sep=1pt, xshift=-0.8em, yshift=0.75em] at (img.north west)
                {{\fontfamily{qhv}\selectfont\bfseries D}};
        \end{tikzpicture}
    \end{subfigure}%
    \caption{
    \textbf{A:} PDF of $z_1$ (blue) and $z_2$ (orange) of the underlying true uniform distribution and $t$-distribution with unit scale estimated by DEC-loss. The two $t$-distributions are highly overlapping whereas there is a clear separation in the uniform distributions. 
    \textbf{B+C} $t$-SNE projection of toy data after training with DEC loss (B) vs our method (C). 
    We first pretrain a simple linear regression model by minimizing an MSE-loss for 25 epochs. Then we are only applying KL loss. \textbf{B:} Visualization of clustering with the DEC loss. Almost all features are assigned to one cluster. No clustered structure visible.  
    \textbf{C:} Clustering of the learned features with our loss. All features get assigned to the right cluster.
     \textbf{D:} Norms of learned cluster centers $\tilde{\mu_1}$ and $\tilde{\mu_2}$ for the DEC-loss. It is clearly visible that the cluster centers collapse after only a few iterations whereas updated cluster centers via DECEMber $\hat{\mu_1}$ and $\hat{\mu_2}$  are converging towards their true mean $\mu_1$ and $\mu_2$, with $\|\mu_1\|_2 = \sqrt{30} \approx 5.48$ and $\| \mu_2\|_2 = \sqrt{30 \cdot 1.5^2} \approx 8.22$.}
    \label{fig:toy_example}
\end{figure}
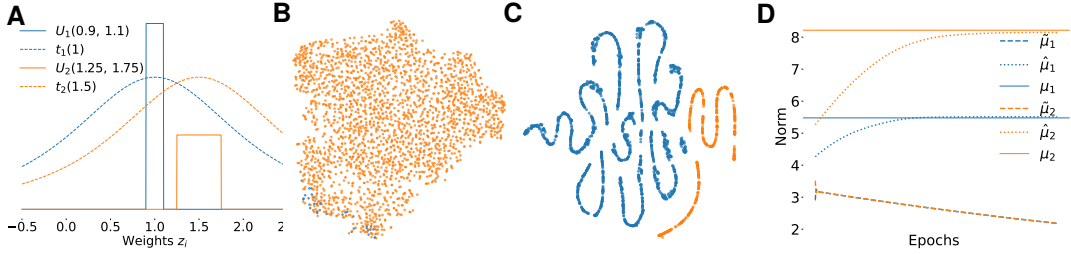

We found that the vanilla DEC loss fails to identify clusters even in this simple toy example, where cluster weights are well-separable after pretraining .
This is because DEC employs a Student’s-$t$ distribution with a fixed unit scale parameter for all clusters, which is too large given how close the two weight distributions of clusters~1 and~2 are (\cref{fig:toy_example}A).
As the magnitude of the weights is given by the regression problem, the scale of the $t$ distribution needs to be adjusted appropriately. 
When this is not done (as in vanilla DEC), the cluster centroids $\tilde{\mu_1}$ and $\tilde{\mu_2}$ rapidly collapse to a single point after only a few iterations (\Cref{fig:toy_example}D).
This happens because there exists a degenerate optimum of the KL divergence: If all cluster centers are equal $\mu_1 = \ldots = \mu_J = c$, plugging them into \cref{eq:qij} 
$q_{ij} = \frac{\left(1 + \lVert z_i - c \rVert^2 \nu^{-1}\right)^{-(\nu + 1)/2}}{\sum_{j'} \left(1 +\lVert z_i - c\rVert^2\nu^{-1}\right)^{-(\nu + 1)/2}} =\frac{1}{J}
$
gives us $p_{ij} = q_{ij}^2 / (\sum_{j'}q_{ij'}^2) = (1/J^2) / (J \cdot (1/J^2)) = 1/J$ as $f_j = f_{j'}$, (\cref{eq:pij}) which means the KL divergence $\mathrm{KL}(P\|Q)=0$, which of course is not a meaningful solution. In DEC, this minimum is not usually found in practice because clusters are initialized with sufficient separation.

To avoid this collapse, we instead used a multivariate $t$-distribution with (diagonal) scale matrices $\Sigma_j$ for each cluster, updating both position and scale with an EM step~(\Cref{alg:model_training}). 
On the same toy example, this approach succeeded in finding good cluster separation (\Cref{fig:toy_example}C), and  the cluster centers $\hat{\mu_1}$ and $\hat{\mu_2}$ converged towards the true underlying locations (\Cref{fig:toy_example}E). 

\paragraph{DECEMber accurately classifies retinal ganglion cells and outperforms conventional clustering approaches.} 
To check whether DECEMber works on real data, we applied it on marmoset retinal ganglion cells (RGCs) where the existence of discrete cell types is well established \cite{masri2019}. 
We used data from two male marmoset retinas published by Sridhar et al.~\cite{sridhar_dataset_2025}, where the neural activity was recorded using a micro-electrode array while presenting grayscale natural movies.

\begin{wrapfigure}[14]{r}{0.5\textwidth}
    \vspace{-12pt}
    \centering
    \begin{subfigure}[t]{0.22\textwidth}
        \begin{tikzpicture}
            \node[anchor=north west, inner sep=0] (img) at (0,0) 
                {\includegraphics[width=\textwidth]{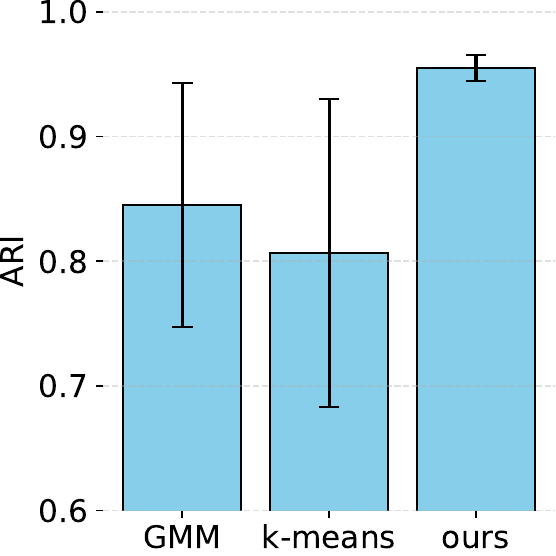}};
            \node[anchor=north west, inner sep=1pt, xshift=0em, yshift=0.8em] at (img.north west)
                {{\fontfamily{qhv}\selectfont\bfseries A}};
        \end{tikzpicture}
    \end{subfigure}%
    \hfill
    \vspace{2mm}
    \begin{subfigure}[t]{0.27\textwidth}
        \begin{tikzpicture}
            \node[anchor=north west, inner sep=0] (img) at (0,0)
                {\includegraphics[width=\textwidth]{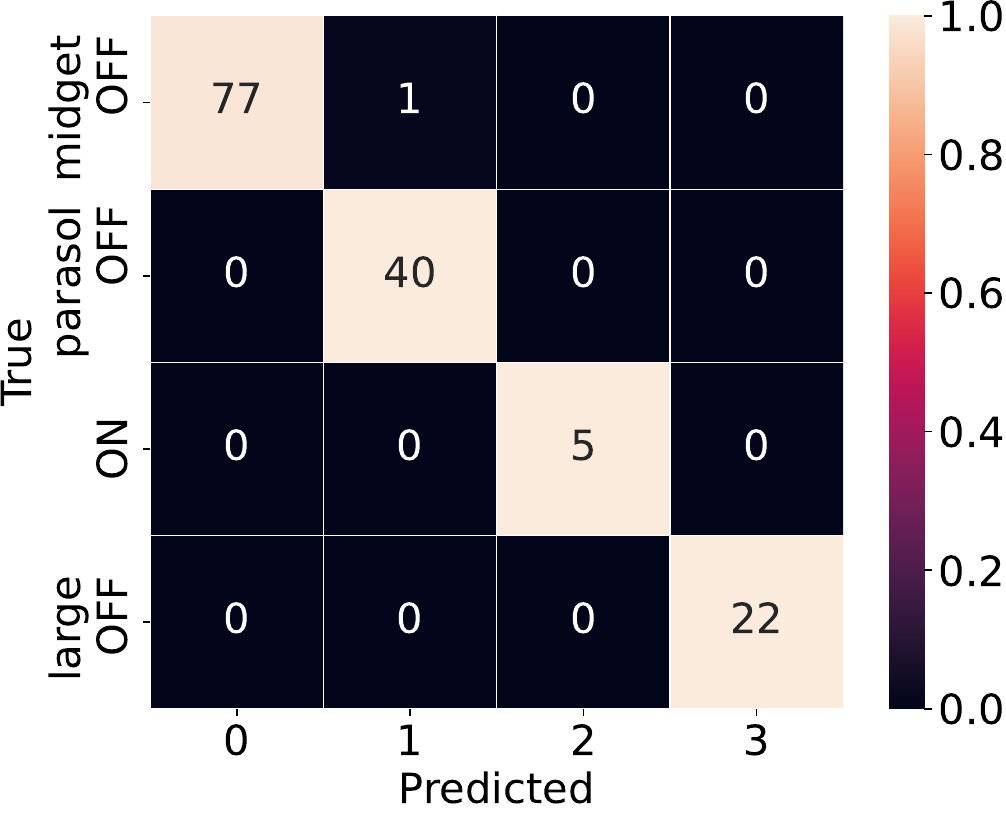}};
            \node[anchor=north west, inner sep=1pt, xshift=0.8em, yshift=0.8em] at (img.north west)
                {{\fontfamily{qhv}\selectfont\bfseries B}};
        \end{tikzpicture}
    \end{subfigure}%
        \caption{
    \textbf{A}: ARI across 3 seeds for GMM, k-means and DECEMber.
    \textbf{B}: DECEMber predictions. Pretraining length: 25 epochs. 
    Corresponding test correlation: $0.805\pm 0.068$ (std).}
    \label{fig:retina_marmoset}
    
\end{wrapfigure}

As we observed substantial differences between the two retinas' temporal response features (potentially due to temperature variation~\cite{zhao2020temporal}), we followed Vystrčilová et al. \cite{Vystrcilova2024} and trained a separate model for each retina to avoid clustering by retina.
We trained the model on all reliably responding cells ($N=235$). However, not all of them corresponded to a known primate RGC type and thus were not assigned a cell type label.
When evaluating DECEMber, we only used the labeled cells.
The first retina contained responses of four cell types (78 midget-OFF-like cells, 40 parasol-OFF-like cells, 5 ON-like cells, and 22 large-OFF cells, further details on classification are in \cref{app:retinaGagnlion}). 

We trained a baseline version of a CNN model \cite{Vystrcilova2024} separately without our proposed clustering loss, using three random seeds. 
Baseline clustering was then performed post hoc using GMM and k-means. Subsequently, we continued training the model with our clustering loss, again using three seeds.

Applied to marmoset retina data, DECEMber achieved reliable classification across cell types, with high clustering consistency (ARI = 0.96 $\pm 0.01$) while maintaining a high predictive performance of 
0.81 $\pm 0.07$.
It surpassed both GMM and k-means, (\Cref{fig:retina_marmoset}A) effectively separating even highly unbalanced groups, such as the ON-cells, resulting in an almost perfect confusion matrix (\Cref{fig:retina_marmoset}B). 
In contrast to k-means, which is sensitive to initialization (Suppl.~\cref{fig:seed_dependence}), DECEMber exhibited greater robustness and aligned more closely with the ground truth labels while the model retained high predictive accuracy. 

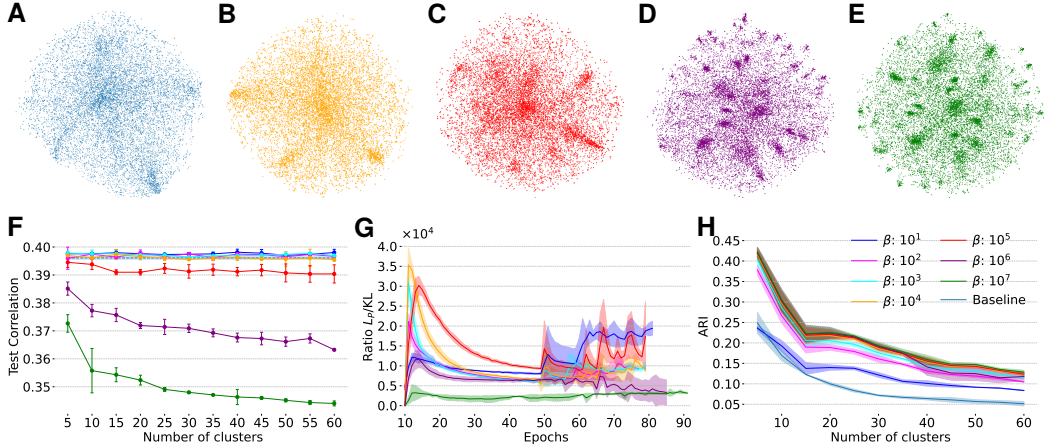
\begin{figure}[t!]
    \begin{subfigure}[t]{0.2\textwidth}
        \begin{tikzpicture}
            \node[anchor=north west, inner sep=0] (img) at (0,0)
                {\includegraphics[width=\textwidth]{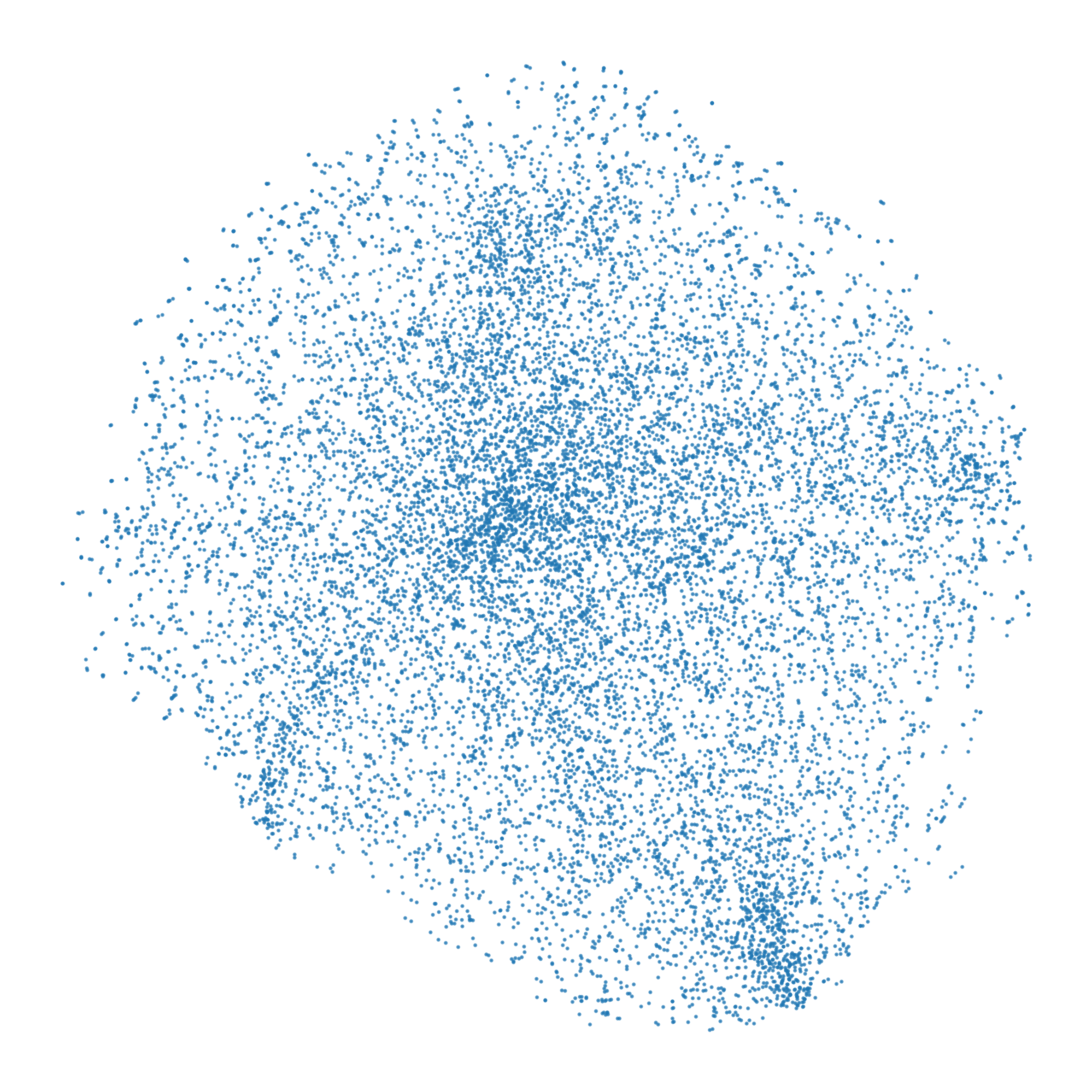}};
            \node[anchor=north west, inner sep=1pt, xshift=0em, yshift=0em] at (img.north west)
                {{\fontfamily{qhv}\selectfont\bfseries A}};
        \end{tikzpicture}
    \end{subfigure}%
    \begin{subfigure}[t]{0.2\textwidth}
        \begin{tikzpicture}
            \node[anchor=north west, inner sep=0] (img) at (0,0)
                {\includegraphics[width=\textwidth]{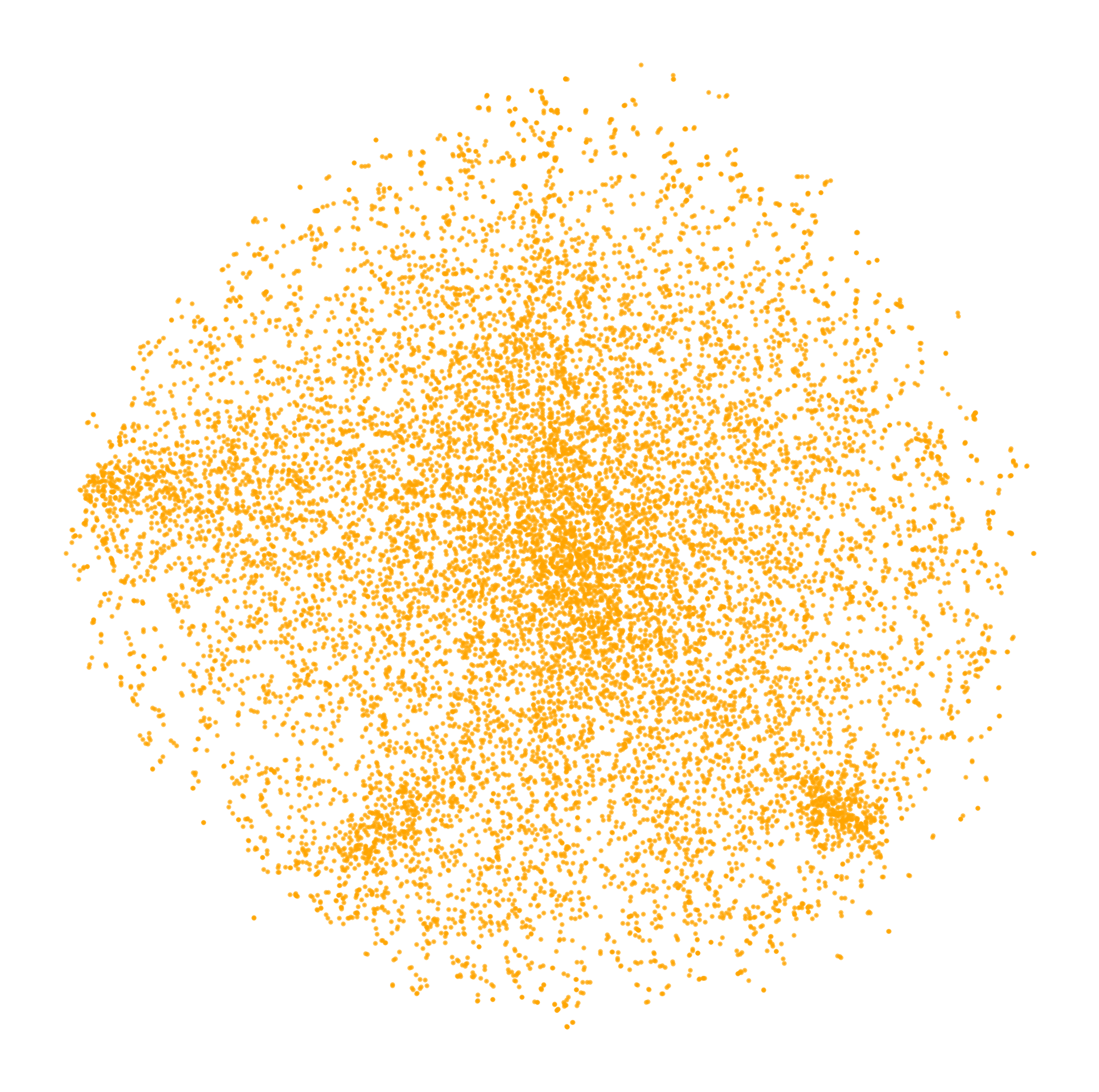}};
            \node[anchor=north west, inner sep=1pt, xshift=0em, yshift=0em] at (img.north west)
                {{\fontfamily{qhv}\selectfont\bfseries B}};
        \end{tikzpicture}
    \end{subfigure}%
    \begin{subfigure}[t]{0.2\textwidth}
        \begin{tikzpicture}
            \node[anchor=north west, inner sep=0] (img) at (0,0)
                {\includegraphics[width=\textwidth]{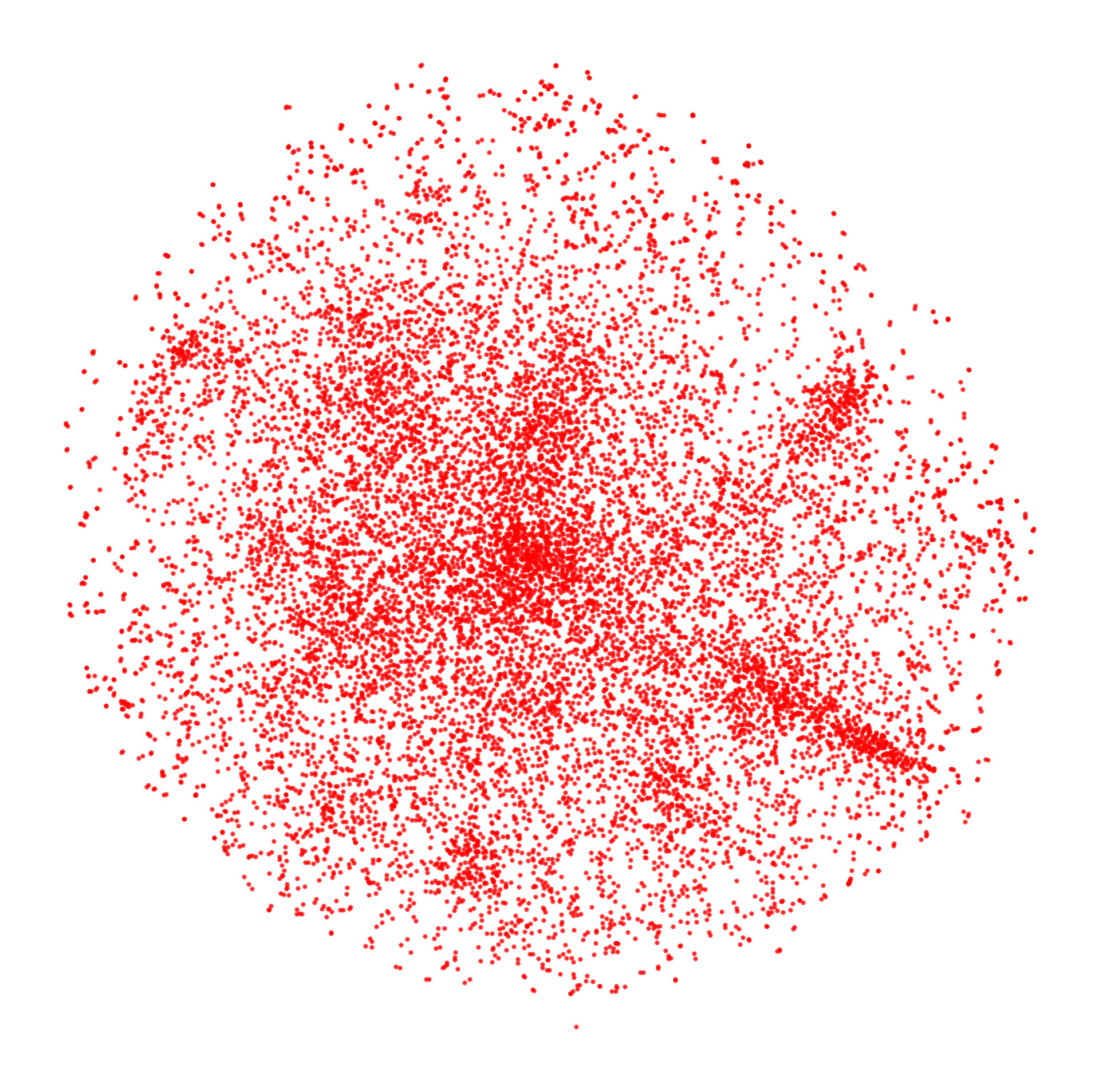}};
            \node[anchor=north west, inner sep=1pt, xshift=0em, yshift=0em] at (img.north west)
                {{\fontfamily{qhv}\selectfont\bfseries C}};
        \end{tikzpicture}
    \end{subfigure}%
    \begin{subfigure}[t]{0.2\textwidth}
        \begin{tikzpicture}
            \node[anchor=north west, inner sep=0] (img) at (0,0)
                {\includegraphics[width=\textwidth]{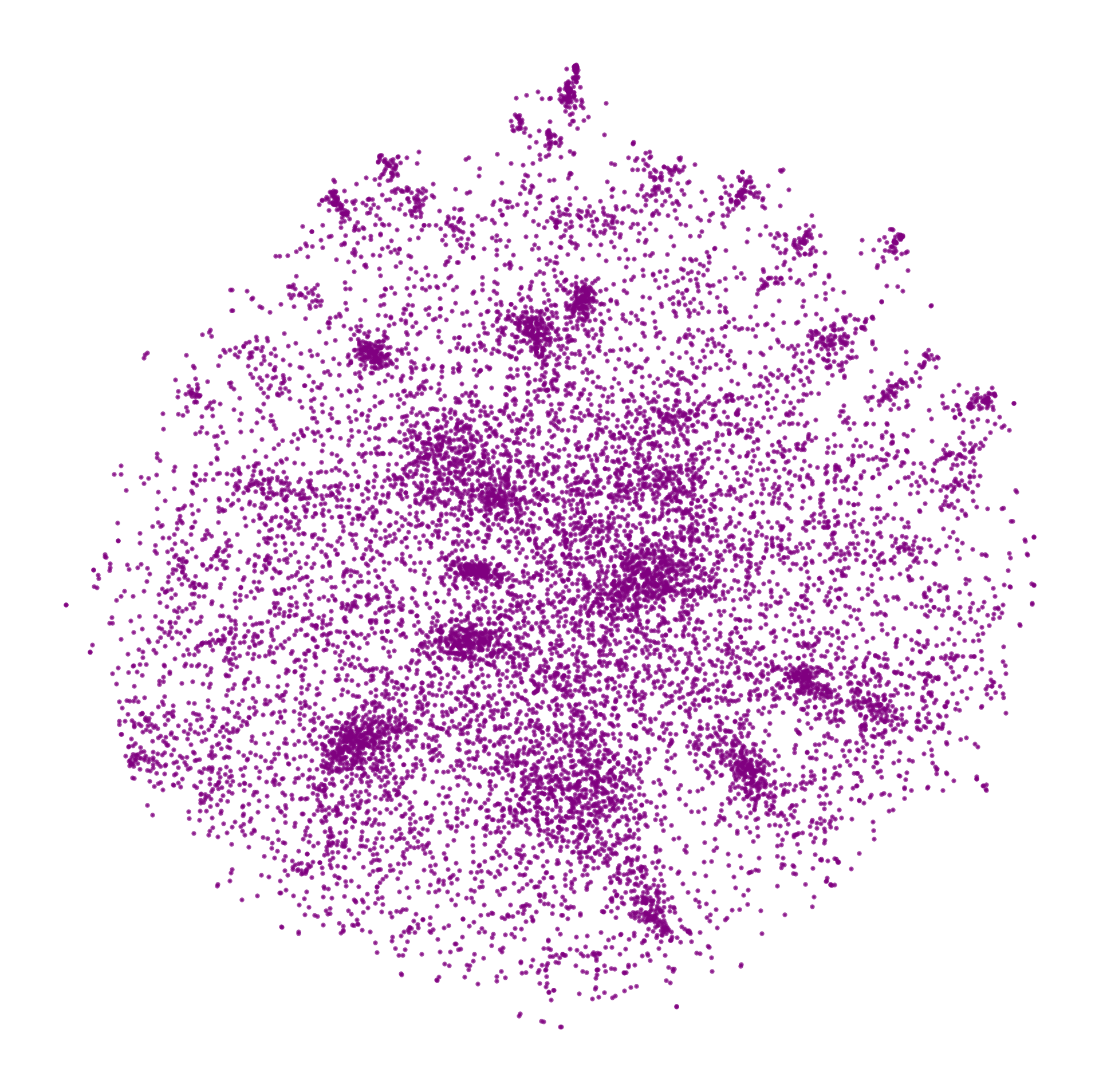}};
            \node[anchor=north west, inner sep=1pt, xshift=0em, yshift=0em] at (img.north west)
                {{\fontfamily{qhv}\selectfont\bfseries D}};
        \end{tikzpicture}
    \end{subfigure}%
    \begin{subfigure}[t]{0.2\textwidth}
        \begin{tikzpicture}
            \node[anchor=north west, inner sep=0] (img) at (0,0)
                {\includegraphics[width=\textwidth]{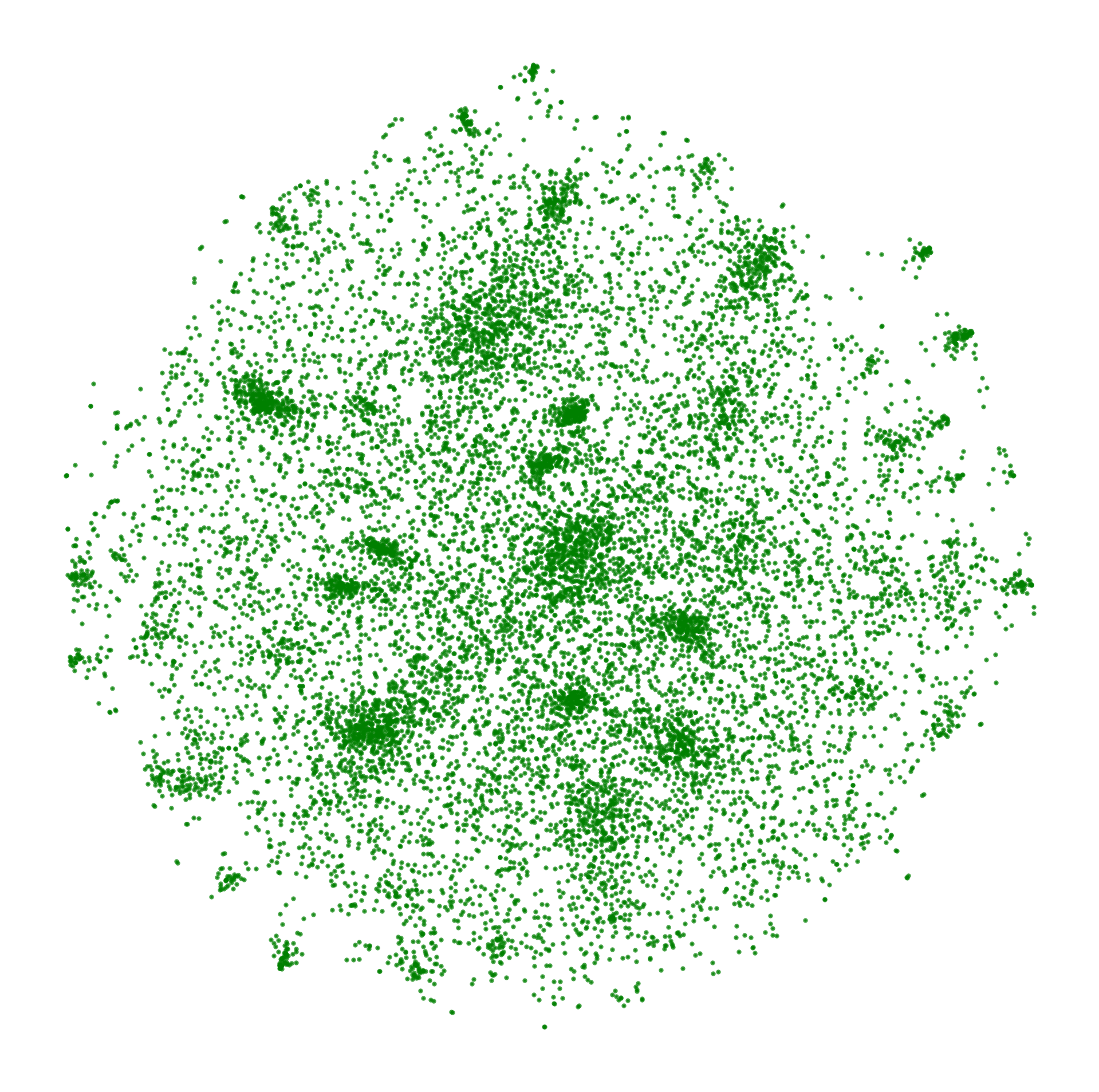}};
            \node[anchor=north west, inner sep=1pt, xshift=0em, yshift=0em] at (img.north west)
                {{\fontfamily{qhv}\selectfont\bfseries E}};
        \end{tikzpicture}
    \end{subfigure}

    \par\vspace{0.1ex}

    \begin{subfigure}[t]{0.33\textwidth}
        \begin{tikzpicture}
            \node[anchor=north west, inner sep=0] (img) at (0,0)
                {\includegraphics[width=\textwidth]{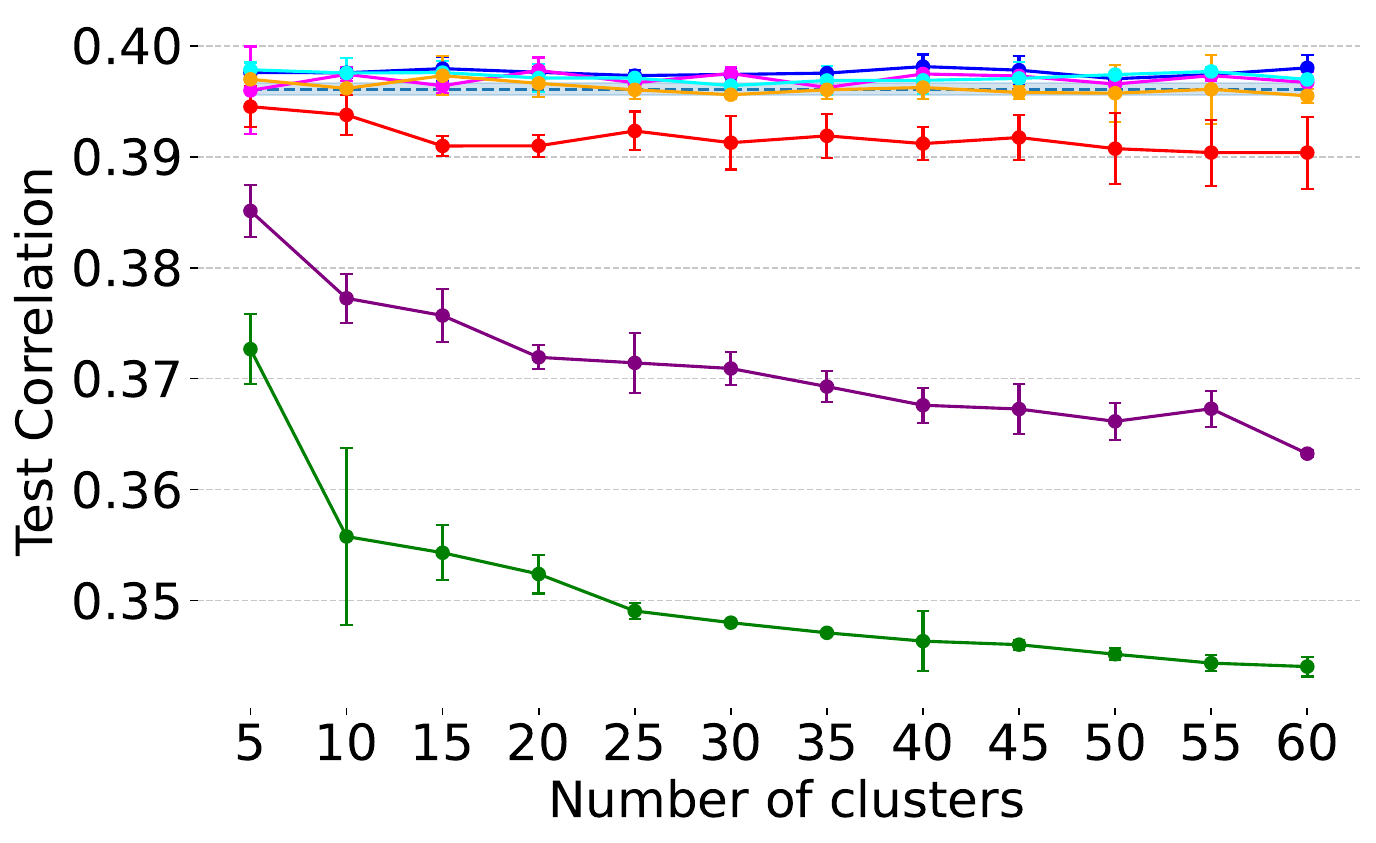}};
            \node[anchor=north west, inner sep=1pt, xshift=0em, yshift=0.8em] at (img.north west)
                {{\fontfamily{qhv}\selectfont\bfseries F}};
        \end{tikzpicture}
    \end{subfigure}%
    \begin{subfigure}[t]{0.32\textwidth}
        \begin{tikzpicture}
            \node[anchor=north west, inner sep=0] (img) at (0,0)
                {\includegraphics[width=\textwidth]{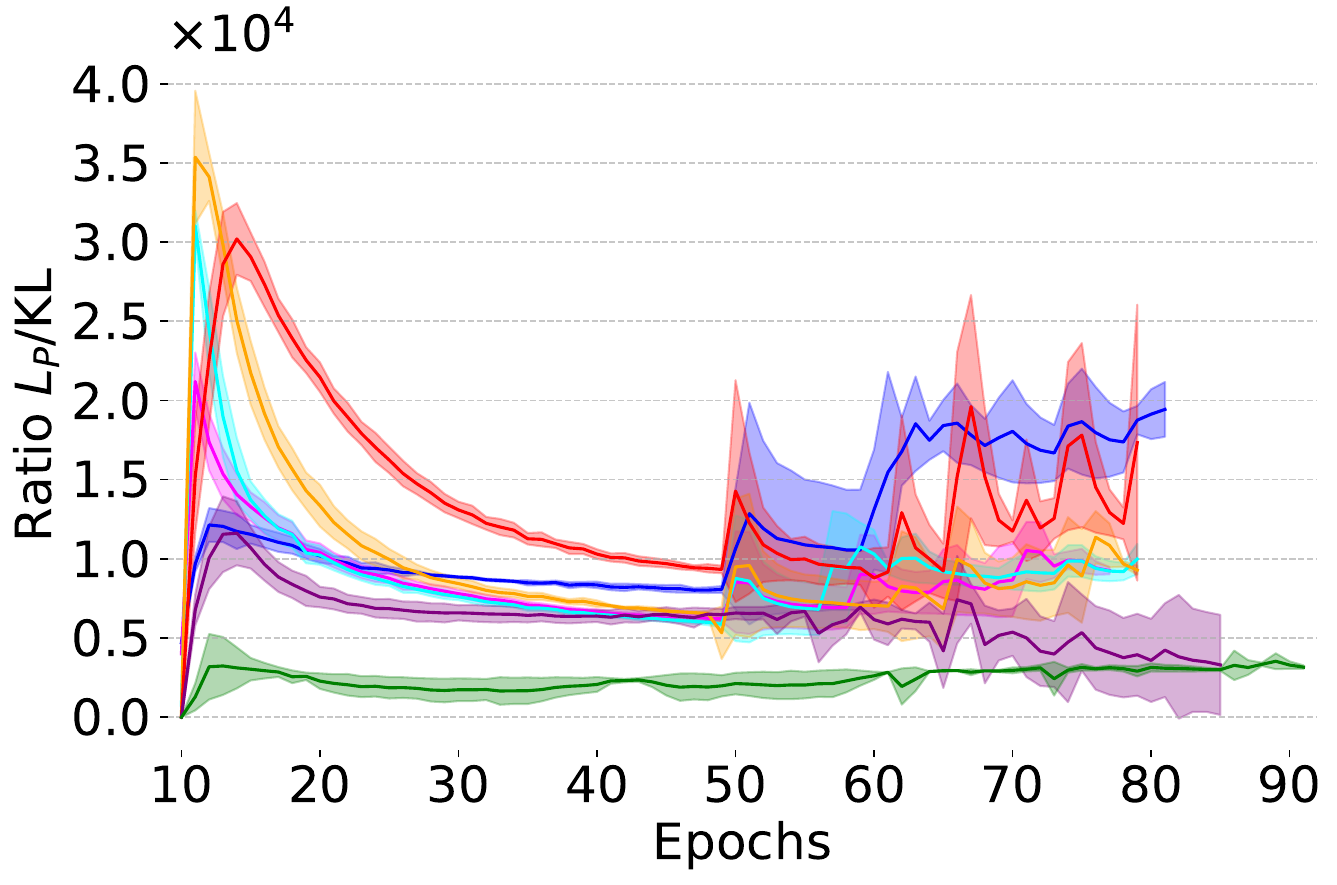}};
            \node[anchor=north west, inner sep=1pt, xshift=-0.3em, yshift=0.4em] at (img.north west)
                {{\fontfamily{qhv}\selectfont\bfseries G}};
        \end{tikzpicture}
    \end{subfigure}
    \begin{subfigure}[t]{0.33\textwidth}
        \begin{tikzpicture}
            \node[anchor=north west, inner sep=0] (img) at (0,0)
                {\includegraphics[width=\textwidth]{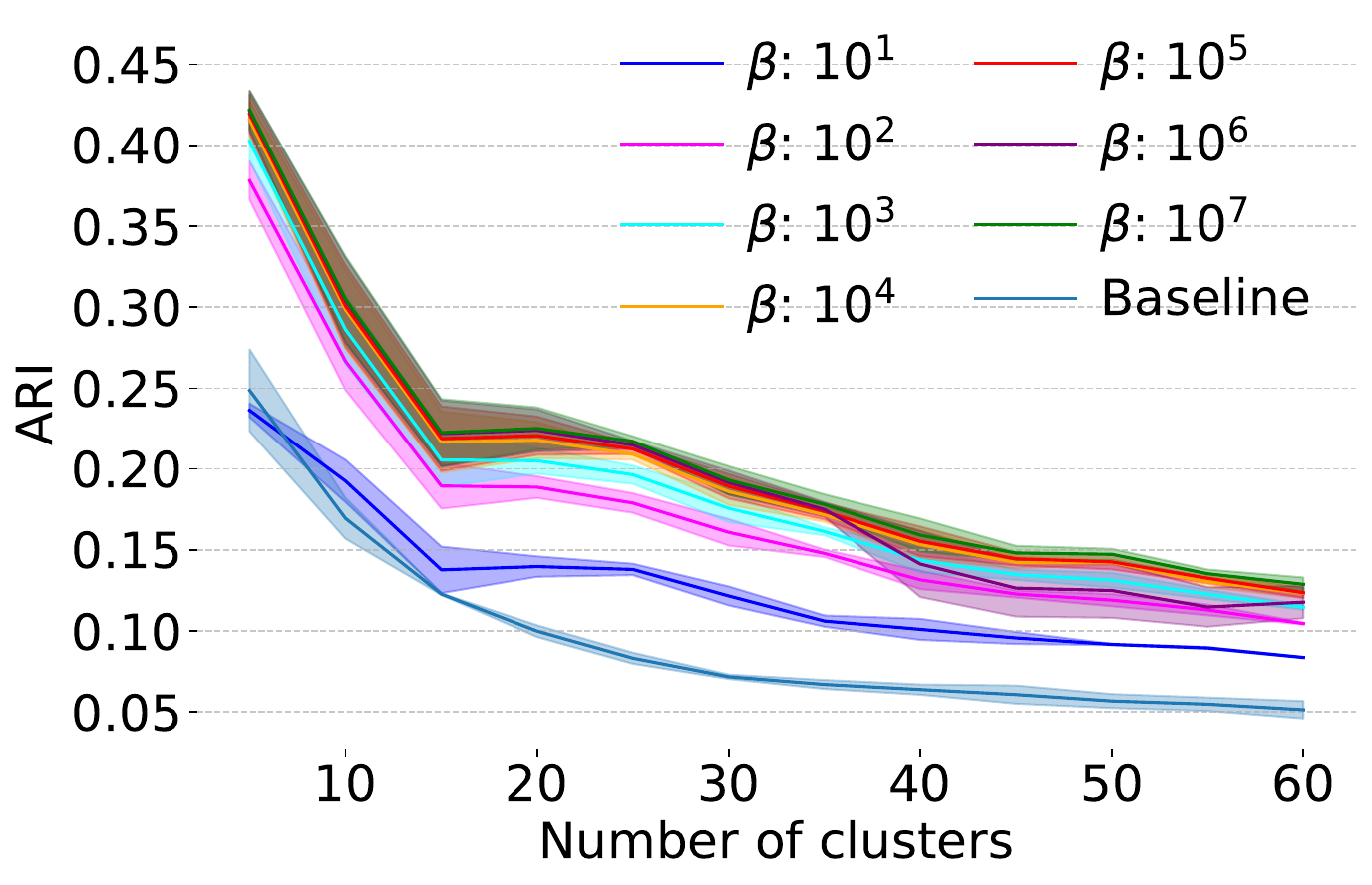}};
            \node[anchor=north west, inner sep=1pt, xshift=0em, yshift=0.4em] at (img.north west)
                {{\fontfamily{qhv}\selectfont\bfseries H}};
        \end{tikzpicture}
    \end{subfigure}

    \caption{\textbf{A}: $t$-SNE of baseline model without clustering loss. \textbf{B-E}: $t$-SNE projections of our model with clustering bias for different multipliers $\beta$ and tuned learning rates (lr). All models use 15 clusters, PE = 10 and seed 100. 
    \textbf{B}: $\beta = 10^4$ and $lr = 0.008$. 
    \textbf{C}: $\beta = 10^5$ and $lr = 0.008$. 
    \textbf{D} $\beta = 10^6$ and $lr = 0.007$. \textbf{E}: $\beta = 10^7$ and $lr = 0.003$. \textbf{F}: Corresponding model performances of the models with clustering bias and differing weights $\beta$, tuned learning rates as described in B-D. \textbf{G}: Ratio of Poisson loss (\cref{L_model} and clustering loss (\cref{eq:KL} for different $\beta$. If $\beta$ is too small $L_{clustering}$ is increasing while $L_p$ is decreasing leading to increasing ratios after epoch 50. All models use PE=10 and 15 clusters.
    \textbf{H}: 
    ARI for different clustering weights $\beta$ with optimal learning rates, PE = 10.}
    \label{fig:qualitative}
\end{figure}

\paragraph{DECEMber enhances local structure among embeddings and hurts performance once it dominates the overall model loss.}
Next, we asked if DECEMber could help to find functional cell types in a visual area without clear known cluster separation.
We used SENSORIUM 2022 dataset and baseline architecture to train a model to predict responses of mouse visual cortex to grayscale images. for seven mice (more detail on data in \cref{app:sensoriumData}). 
Previous work~\cite{turishcheva202, ivan} observed density modes in the functional embeddings of mouse V1 neurons (\Cref{fig:qualitative}A) and hypothesized that these modes may correspond to discrete functional cell types. 
To investigate whether these patterns reflect true discrete and distinct cell types, we applied DECEMber (\Cref{alg:model_training}), hypothesizing that if such types exist, DECEMber would help to separate them.
As the number of excitatory cell types in the mouse visual cortex remains unclear, with estimates ranging from 20 to 50 \cite{gouwens2019classification, ivan}, we considered a range of $j=5, ..., 60$ in increments of 5.

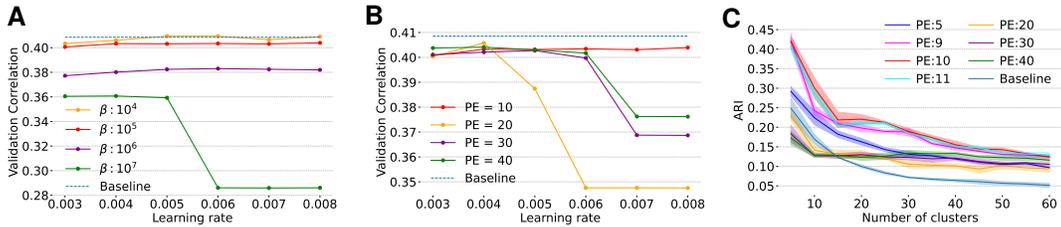
\begin{figure}[t!]
    \centering
    \begin{subfigure}[t]{0.32\textwidth}
        \begin{tikzpicture}
            \node[anchor=north west, inner sep=0] (img) at (0,0) 
                {\includegraphics[width=\textwidth]{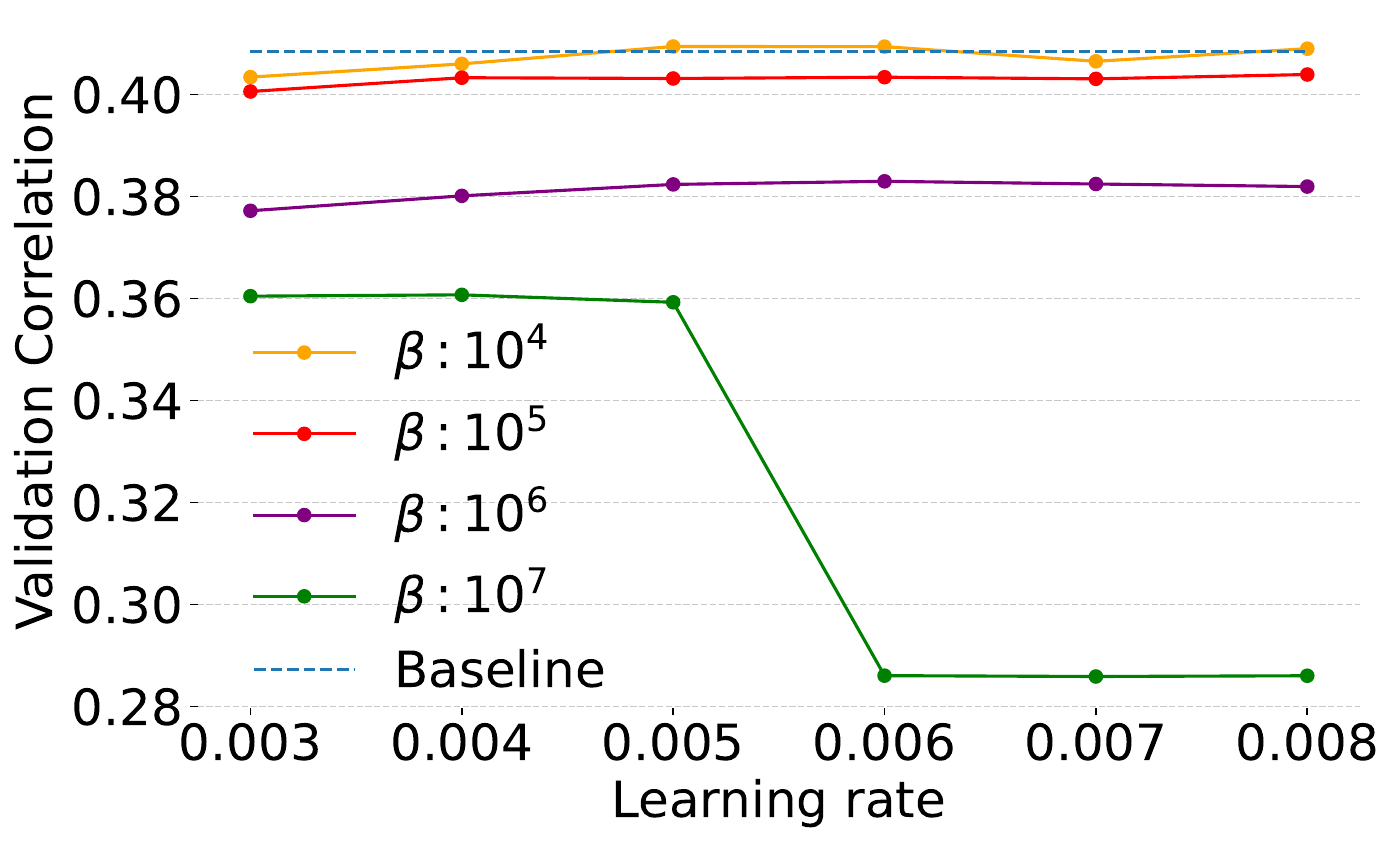}};
            \node[anchor=north west, inner sep=1pt, xshift=0em, yshift=0.8em] at (img.north west)
                {{\fontfamily{qhv}\selectfont\bfseries A}};
        \end{tikzpicture}
    \end{subfigure}%
    \hfill
    \begin{subfigure}[t]{0.32\textwidth}
        \begin{tikzpicture}
            \node[anchor=north west, inner sep=0] (img) at (0,0)
                {\includegraphics[width=\textwidth]{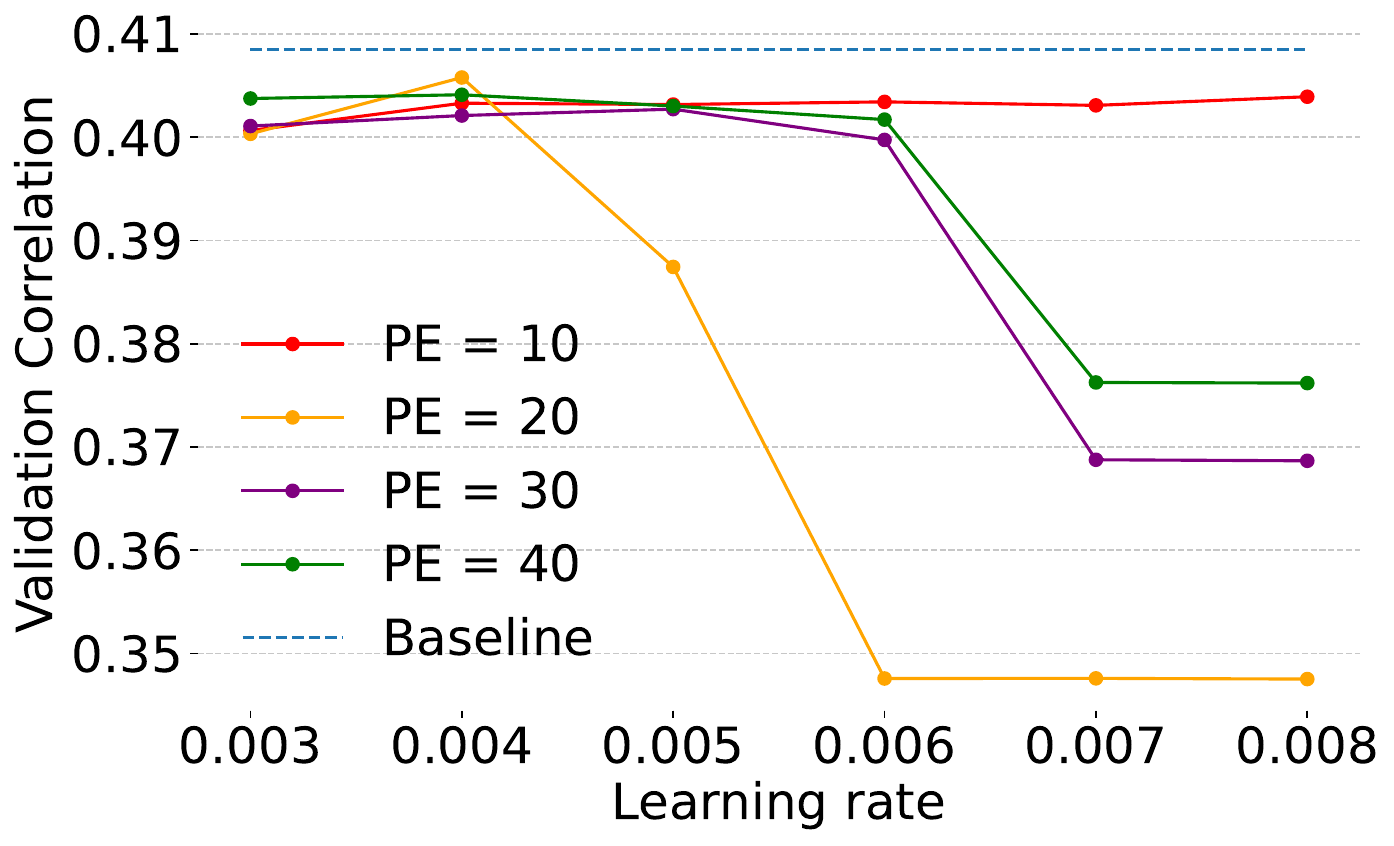}};
            \node[anchor=north west, inner sep=1pt, xshift=-0.4em, yshift=0.8em] at (img.north west)
                {{\fontfamily{qhv}\selectfont\bfseries B}};
        \end{tikzpicture}
    \end{subfigure}%
    \hfill
    \begin{subfigure}[t]{0.32\textwidth}
        \begin{tikzpicture}
            \node[anchor=north west, inner sep=0] (img) at (0,0)
                {\includegraphics[width=\textwidth]{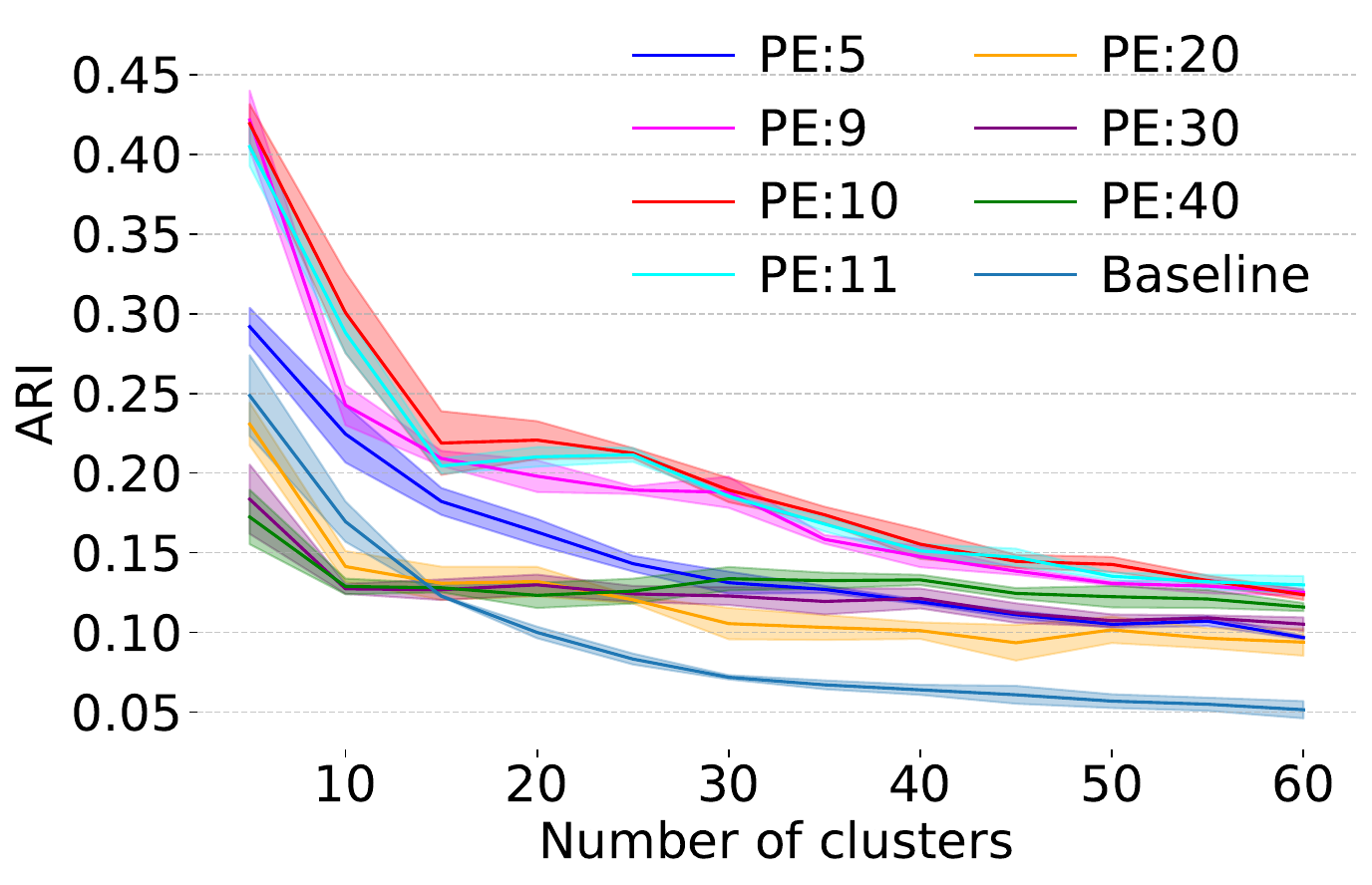}};
            \node[anchor=north west, inner sep=1pt, xshift=-0.4em, yshift=0.2em] at (img.north west)
                {{\fontfamily{qhv}\selectfont\bfseries C}};
        \end{tikzpicture}
    \end{subfigure}
    \caption{
    \textbf{A--B}: Learning rate tuning for $\beta$ (A) and length of pretraining (B). We fixed amount of clusters to 15. If the learning rate is too high the clustering loss starts oscillating due to learning rate scheduling leading to a massive drop in performance. 
    \textbf{C} ARI for different number of pretraining epochs vs baseline model. 
    For each number of pretraining epochs we used the optimal learning rate and set $\beta=10^5$ for all experiments. 
    All settings of DECEMber have better cluster structures after 15 clusters at latest. 
    It is visible that 10 pretraining epochs generate the best clustered embeddings.}
    \label{fig:quantitative}
\end{figure}

We tested a wide range of loss strengths $\beta$, to find the optimal value for this parameter.
As $\beta$ increases, t-SNE vizualization suggests improved qualitative separation of clusters in the embedding space (\Cref{fig:qualitative}B–E). 
However, this comes at a cost: when the clustering loss becomes dominant, the model’s predictive performance drops significantly (\Cref{fig:qualitative}F). 
This made us question if the qualitative structure in t-SNE plots is meaningful.
To answer this question, we quantified clustering consistency using ARI between three model fits with different seeds and found that clustering consistency noticeably improved compared to the GMM baseline.
We see the ARI improvement as long as beta does not hurt performance ($\beta \leq 10^4$; \Cref{fig:qualitative}H, dark blue, magenta, cyan and yellow lines).
However, once $\beta > 10^4$, performance starts suffering (\Cref{fig:qualitative}F) and the ARI does not improve anymore (\Cref{fig:qualitative}H), suggesting that the structure is created by removing functionally relevant heterogeneity between neurons. 
While ARI values double compared to the baseline model, there is no clear peak around a certain number of clusters. We would expect ARI to peak noticeably at the ``true'' number of clusters if such a structure existed. 
This suggests that mouse V1 likely lacks discrete functional cell types. 
Still, the clear improvement indicates meaningful local structure in functional embeddings.


\begin{wrapfigure}{r}{0.32\textwidth}
    \vspace{-20pt} 
    \centering
    \includegraphics[width=0.32\textwidth]{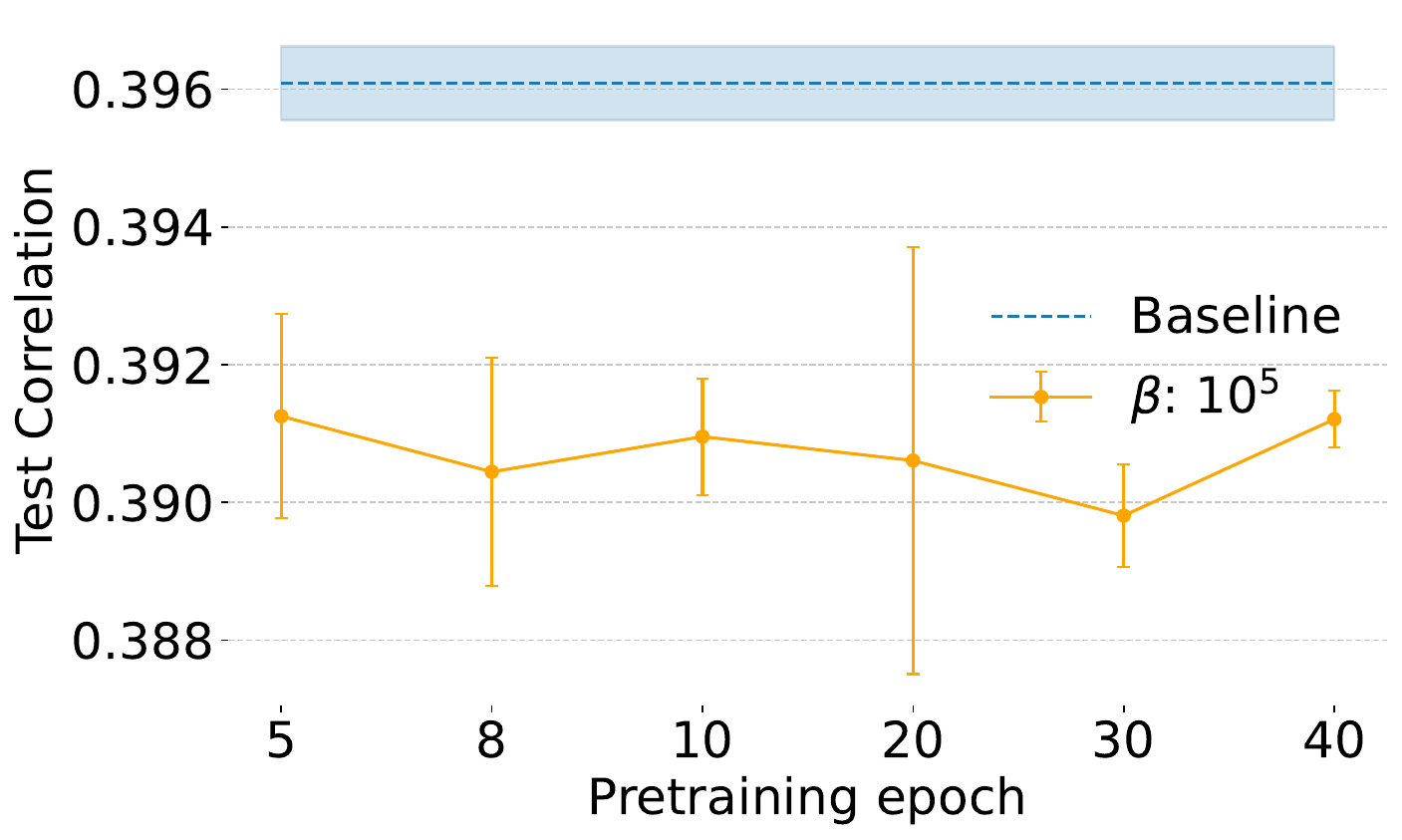}
    \caption{The choice of pretrain epoch doesn't influence performance when we're using an optimal learning rate. }\label{fig:test_correlation_pe}
    \vspace{-1.2em} 
\end{wrapfigure}

\paragraph{Consistency of embeddings depends on length of pretraining.}
To validate our conclusions that mouse V1 lacks discrete functional cell types, we performed extensive tuning of DECEMber by using different numbers of pretraining epochs before turning on the clustering loss, different clustering strengths $\beta$ and tuned learning rates to optimize model predictive performance.  
Across all settings, DECEMber achieved higher ARI scores than the baseline, indicating better consistency of embeddings (\Cref{fig:qualitative}H, \Cref{fig:quantitative}C).

We found that the optimal learning rate varied depending on the number of pretraining epochs (\Cref{fig:quantitative}A), and also depended on the clustering loss strength $\beta$ (\Cref{fig:quantitative}B).
Importantly, the choice of the number of pretraining epochs had minimal effect on the overall predictive performance if the learning rate was optimally chosen, with differences staying within the standard deviation across runs. We tuned on the validation set (\Cref{fig:quantitative}B), and checked on the test set (\Cref{fig:test_correlation_pe}). 
However, we observed a distinct peak in ARI  for 10 pretraining epochs in the case of the mouse visual cortex (\Cref{fig:quantitative}C).
While the ARI improved across a range of cluster settings, we did not observe a sharp maximum at any specific cluster count.



\paragraph{DECEMber improves embeddings across different datasets and model architectures.}
\label{sec:generalization}

To ensure that DECEMber generalizes across architectures, modalities, and species, we additionally tested it on data from the mouse retina and macaque visual cortex area V4.
We did not extensively tune hyperparameters, we only decreased the learning rate (lr) to stabilize the baseline model training for both datasets, and set $\beta$ as described in \Cref{sec:beta} (exact settings in \Cref{app:expSetting}).
More extensive tuning of the lr, $\beta$ or the number of pretraining epochs can lead to better results. 
For both datasets we preserved the performance of the original models (\Cref{app:additionalPerformances}).

For the mouse retina we used both the data and the models from Höfling et al. \cite{hofling2024chromatic}.
As for the marmoset retina, we trained a separate model for each retina to account for the temperature differences between retinas.
Given the limited availability of cell-type labels, we included all cells in our analysis and evaluated cluster consistency across varying numbers of clusters. For details on dataset averaging and per-dataset analysis, see \Cref{app:lara}.

\begin{wrapfigure}{r}{0.66\textwidth}
    \vspace{-12pt} 
    \centering
    \includegraphics[width=0.66\textwidth]{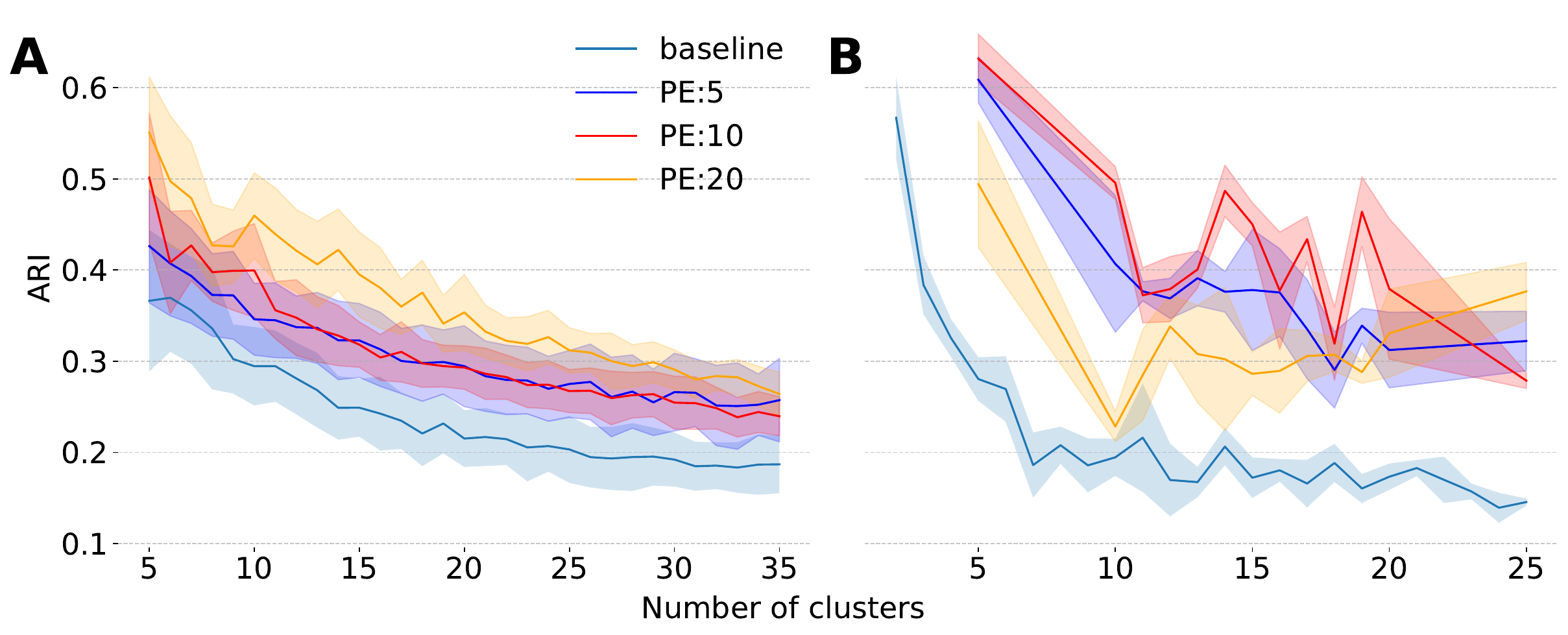}
    \caption{
    ARI on
    \textbf{A} mouse retina \cite{hofling2024chromatic}, weighted across six models.
    \textbf{B} monkeys V4 \cite{willeke2023deep}.
    } 
    \label{fig:monkey}
    \vspace{-1.2em}
\end{wrapfigure}

For macaque V4 data we used spiking extracellular multi-electrode recorded responses of neurons to gray-scale natural images shown to awake macaque monkeys \cite{willeke2023deep} and the model from  Pierzchlewicz et al. \cite{pierzchlewicz2023energy}, which had a different readout architecture -- an attention readout instead of the previously used Gaussian readout.
We trained the model on 1000 cells and measured the ARI across three model seeds. 
The embedding consistency doubled using our method (\Cref{fig:monkey}B).
This shows that DECEMber is robust not only across different data modalities but also across architectures.
For more analysis of the monkey dataset see \Cref{app:monkeys}.

\section{Discussion}
\label{sec:discussion}
In this work we introduced DECEMber, an additional training loss with explicit clustering bias for predictive models of neuronal responses.
DECEMber enhances cluster consistency, while keeping state-of-the-art predictive performance.
It is robust across different data modalities (electrophysiology and calcium imaging), species (mice, primates) and visual areas (retina, V1, V4).
We also showed that DECEMber is robust across both static (mouse V1, macaque V4) and dynamic (retinas) cores and multiple readout architectures -- the Gaussian readout and the attention readout.

We see DECEMber as a model-driven hypothesis test: if clear functional cell types exist, then incorporating this bias should improve model performance and/or the embedding structure, which we measure as cluster consistency.
While improvements are observed across datasets and architectures, our main focus was mouse V1, where the existence of discrete excitatory cell types remains debated.
Our results support the idea that excitatory neurons in mouse visual cortex form a functional continuum rather than discrete clusters.
This finding is consistent with recent work studying different modalities by Weiler et al. \cite{Weiler2023}, Tong et al. \cite{Tong2023.11.03.565500}, and Weis et al. \cite{Weis2025unsupervised}, who independently found no clear boundaries in morphological or electrophysiological features. 
In line with Zeng \cite{zeng2022cell}, we argue that future efforts to define mouse V1 cell types should emphasize multi-modality combining functional, morphological, and genetic data. 
This approach has proven fruitful in the retina, where functional types alone are coarser than those derived from multiple modalities \cite{baden2016functional, burg2024discriminativestimulifunctionalcell}.


Given the generality of our clustering loss, which is model-agnostic and not tied to a specific architecture, we believe it holds promise for use in multi-modal models aiming to define cell types or in broader unsupervised representation learning contexts. 

\paragraph{Limitations} 
DECEMber requires a predefined number of clusters. 
When this is unknown, multiple runs with varying cluster counts and seeds are necessary in combination with an evaluation ARI-like metric to identify the optimal configuration. 
This increases the computational cost. 
Choosing an appropriate clustering strength $\beta$ is also crucial for balancing ARI and model performance and further work is needed to determine the optimal pretraining duration.

Moreover, operating in high-dimensional feature spaces introduces an additional challenge: the cluster covariance matrices can become large and ill-conditioned, with tiny diagonal values, hitting the limits of numerical stability.
We address this issue by clamping small values, though this solution is heuristic rather than principled. 
Furthermore, high-dimensional settings require a sufficient number of data points to prevent overfitting of the scale matrices. 
Additionally, we currently assume a $t$-distributed feature space via a $t$-mixture model, but this can be adjusted if a more suitable prior over the embeddings is known.





\newpage
\bibliographystyle{unsrtnat}
\bibliography{bibliography} 

\begin{thebibliography}{56}
\providecommand{\natexlab}[1]{#1}
\providecommand{\url}[1]{\texttt{#1}}
\expandafter\ifx\csname urlstyle\endcsname\relax
  \providecommand{\doi}[1]{doi: #1}\else
  \providecommand{\doi}{doi: \begingroup \urlstyle{rm}\Url}\fi

\bibitem[Zeng(2022)]{zeng2022cell}
Hongkui Zeng.
\newblock What is a cell type and how to define it?
\newblock \emph{Cell}, 185\penalty0 (15):\penalty0 2739--2755, 2022.

\bibitem[DeFelipe et~al.(2013)DeFelipe, L{\'o}pez-Cruz, Benavides-Piccione,
  Bielza, Larra{\~n}aga, Anderson, Burkhalter, Cauli, Fair{\'e}n, Feldmeyer,
  et~al.]{defelipe2013new}
Javier DeFelipe, Pedro~L L{\'o}pez-Cruz, Ruth Benavides-Piccione, Concha
  Bielza, Pedro Larra{\~n}aga, Stewart Anderson, Andreas Burkhalter, Bruno
  Cauli, Alfonso Fair{\'e}n, Dirk Feldmeyer, et~al.
\newblock New insights into the classification and nomenclature of cortical
  gabaergic interneurons.
\newblock \emph{Nature Reviews Neuroscience}, 14\penalty0 (3):\penalty0
  202--216, 2013.

\bibitem[Oberlaender et~al.(2012)Oberlaender, De~Kock, Bruno, Ramirez, Meyer,
  Dercksen, Helmstaedter, and Sakmann]{oberlaender2012cell}
Marcel Oberlaender, Christiaan~PJ De~Kock, Randy~M Bruno, Alejandro Ramirez,
  Hanno~S Meyer, Vincent~J Dercksen, Moritz Helmstaedter, and Bert Sakmann.
\newblock Cell type--specific three-dimensional structure of thalamocortical
  circuits in a column of rat vibrissal cortex.
\newblock \emph{Cerebral cortex}, 22\penalty0 (10):\penalty0 2375--2391, 2012.

\bibitem[Markram et~al.(2015)Markram, Muller, Ramaswamy, Reimann, Abdellah,
  Sanchez, Ailamaki, Alonso-Nanclares, Antille, Arsever,
  et~al.]{markram2015reconstruction}
Henry Markram, Eilif Muller, Srikanth Ramaswamy, Michael~W Reimann, Marwan
  Abdellah, Carlos~Aguado Sanchez, Anastasia Ailamaki, Lidia Alonso-Nanclares,
  Nicolas Antille, Selim Arsever, et~al.
\newblock Reconstruction and simulation of neocortical microcircuitry.
\newblock \emph{Cell}, 163\penalty0 (2):\penalty0 456--492, 2015.

\bibitem[Scala et~al.(2019)Scala, Kobak, Shan, Bernaerts, Laturnus, Cadwell,
  Hartmanis, Froudarakis, Castro, Tan, et~al.]{scala2019layer}
Federico Scala, Dmitry Kobak, Shen Shan, Yves Bernaerts, Sophie Laturnus,
  Cathryn~Rene Cadwell, Leonard Hartmanis, Emmanouil Froudarakis, Jesus~Ramon
  Castro, Zheng~Huan Tan, et~al.
\newblock Layer 4 of mouse neocortex differs in cell types and circuit
  organization between sensory areas.
\newblock \emph{Nature communications}, 10\penalty0 (1):\penalty0 4174, 2019.

\bibitem[Weis et~al.(2025)Weis, Papadopoulos, Hansel, L{\"u}ddecke, Celii,
  Fahey, Wang, Bae, Bodor, Brittain, Buchanan, Bumbarger, Castro, Collman, {da
  Costa}, Dorkenwald, Elabbady, Halageri, Jia, Jordan, Kapner, Kemnitz, Kinn,
  Lee, Li, Lu, Macrina, Mahalingam, Mitchell, Mondal, Mu, Nehoran, Popovych,
  Reid, {Schneider-Mizell}, Seung, Silversmith, Takeno, Torres, Turner, Wong,
  Wu, Yin, Yu, Reimer, Berens, Tolias, and Ecker]{Weis2025unsupervised}
Marissa~A. Weis, Stelios Papadopoulos, Laura Hansel, Timo L{\"u}ddecke, Brendan
  Celii, Paul~G. Fahey, Eric~Y. Wang, J.~Alexander Bae, Agnes~L. Bodor, Derrick
  Brittain, JoAnn Buchanan, Daniel~J. Bumbarger, Manuel~A. Castro, Forrest
  Collman, Nuno~Ma{\c c}arico {da Costa}, Sven Dorkenwald, Leila Elabbady,
  Akhilesh Halageri, Zhen Jia, Chris Jordan, Dan Kapner, Nico Kemnitz, Sam
  Kinn, Kisuk Lee, Kai Li, Ran Lu, Thomas Macrina, Gayathri Mahalingam, Eric
  Mitchell, Shanka~Subhra Mondal, Shang Mu, Barak Nehoran, Sergiy Popovych,
  R.~Clay Reid, Casey~M. {Schneider-Mizell}, H.~Sebastian Seung, William
  Silversmith, Marc Takeno, Russel Torres, Nicholas~L. Turner, William Wong,
  Jingpeng Wu, Wenjing Yin, Szi-chieh Yu, Jacob Reimer, Philipp Berens,
  Andreas~S. Tolias, and Alexander~S. Ecker.
\newblock An unsupervised map of excitatory neuron dendritic morphology in the
  mouse visual cortex.
\newblock \emph{Nature Communications}, 16\penalty0 (1):\penalty0 3361, April
  2025.
\newblock ISSN 2041-1723.
\newblock \doi{10.1038/s41467-025-58763-w}.

\bibitem[Baden et~al.(2016)Baden, Berens, Franke, Rom{\'a}n~Ros{\'o}n, Bethge,
  and Euler]{baden2016functional}
Tom Baden, Philipp Berens, Katrin Franke, Miroslav Rom{\'a}n~Ros{\'o}n,
  Matthias Bethge, and Thomas Euler.
\newblock The functional diversity of retinal ganglion cells in the mouse.
\newblock \emph{Nature}, 529\penalty0 (7586):\penalty0 345--350, 2016.

\bibitem[Willeke et~al.(2022)Willeke, Fahey, Bashiri, Pede, Burg, Blessing,
  Cadena, Ding, Lurz, Ponder, Muhammad, Patel, Ecker, Tolias, and
  Sinz]{sensorium}
Konstantin~F. Willeke, Paul~G. Fahey, Mohammad Bashiri, Laura Pede, Max~F.
  Burg, Christoph Blessing, Santiago~A. Cadena, Zhiwei Ding, Konstantin-Klemens
  Lurz, Kayla Ponder, Taliah Muhammad, Saumil~S. Patel, Alexander~S. Ecker,
  Andreas~S. Tolias, and Fabian~H. Sinz.
\newblock The sensorium competition on predicting large-scale mouse primary
  visual cortex activity, 2022.
\newblock URL \url{https://arxiv.org/abs/2206.08666}.

\bibitem[Turishcheva et~al.(2024{\natexlab{a}})Turishcheva, Fahey,
  Vystr{\v{c}}ilov{\'a}, Hansel, Froebe, Ponder, Qiu, Willeke, Bashiri, Wang,
  et~al.]{turishcheva2024dynamic}
Polina Turishcheva, Paul~G Fahey, Michaela Vystr{\v{c}}ilov{\'a}, Laura Hansel,
  Rachel Froebe, Kayla Ponder, Yongrong Qiu, Konstantin~F Willeke, Mohammad
  Bashiri, Eric Wang, et~al.
\newblock The dynamic sensorium competition for predicting large-scale mouse
  visual cortex activity from videos.
\newblock \emph{ArXiv}, pages arXiv--2305, 2024{\natexlab{a}}.

\bibitem[Antolík et~al.(2016)Antolík, Hofer, Bednar, and
  Mrsic-Flogel]{Antolik}
Ján Antolík, Sonja~B. Hofer, James~A. Bednar, and Thomas~D. Mrsic-Flogel.
\newblock Model constrained by visual hierarchy improves prediction of neural
  responses to natural scenes.
\newblock \emph{PLOS Computational Biology}, 12\penalty0 (6):\penalty0 1--22,
  06 2016.
\newblock \doi{10.1371/journal.pcbi.1004927}.
\newblock URL \url{https://doi.org/10.1371/journal.pcbi.1004927}.

\bibitem[Schmidt et~al.(2025)Schmidt, Turishcheva, Shrinivasan, and
  Sinz]{schmidt2025modelingdynamicneuralactivity}
Finn Schmidt, Polina Turishcheva, Suhas Shrinivasan, and Fabian~H. Sinz.
\newblock Modeling dynamic neural activity by combining naturalistic video
  stimuli and stimulus-independent latent factors, 2025.
\newblock URL \url{https://arxiv.org/abs/2410.16136}.

\bibitem[Wang et~al.(2023)Wang, Fahey, Ponder, Ding, Chang, Muhammad, Patel,
  Ding, Tran, Fu, Papadopoulos, Franke, Ecker, Reimer, Pitkow, Sinz, and
  Tolias]{Wang2023.03.21.533548}
Eric~Y. Wang, Paul~G. Fahey, Kayla Ponder, Zhuokun Ding, Andersen Chang, Taliah
  Muhammad, Saumil Patel, Zhiwei Ding, Dat Tran, Jiakun Fu, Stelios
  Papadopoulos, Katrin Franke, Alexander~S. Ecker, Jacob Reimer, Xaq Pitkow,
  Fabian~H. Sinz, and Andreas~S. Tolias.
\newblock Towards a foundation model of the mouse visual cortex.
\newblock \emph{bioRxiv}, 2023.
\newblock \doi{10.1101/2023.03.21.533548}.
\newblock URL
  \url{https://www.biorxiv.org/content/early/2023/03/24/2023.03.21.533548}.

\bibitem[Turishcheva et~al.(2024{\natexlab{b}})Turishcheva, Burg, Sinz, and
  Ecker]{turishcheva202}
Polina Turishcheva, Max Burg, Fabian~H. Sinz, and Alexander Ecker.
\newblock Reproducibility of predictive networks for mouse visual cortex,
  2024{\natexlab{b}}.
\newblock URL \url{https://arxiv.org/abs/2406.12625}.

\bibitem[Franke et~al.(2021)Franke, Willeke, Ponder, Galdamez, Muhammad, Patel,
  Froudarakis, Reimer, Sinz, and Tolias]{franke2021behavioral}
Katrin Franke, Konstantin~F Willeke, Kayla Ponder, Mario Galdamez, Taliah
  Muhammad, Saumil Patel, Emmanouil Froudarakis, Jacob Reimer, Fabian Sinz, and
  Andreas~S Tolias.
\newblock Behavioral state tunes mouse vision to ethological features through
  pupil dilation.
\newblock \emph{bioRxiv}, pages 2021--09, 2021.

\bibitem[Ecker et~al.(2018)Ecker, Sinz, Froudarakis, Fahey, Cadena, Walker,
  Cobos, Reimer, Tolias, and Bethge]{ecker2018rotation}
Alexander~S Ecker, Fabian~H Sinz, Emmanouil Froudarakis, Paul~G Fahey,
  Santiago~A Cadena, Edgar~Y Walker, Erick Cobos, Jacob Reimer, Andreas~S
  Tolias, and Matthias Bethge.
\newblock A rotation-equivariant convolutional neural network model of primary
  visual cortex.
\newblock \emph{arXiv preprint arXiv:1809.10504}, 2018.

\bibitem[Vystr{\v c}ilov{\'a} et~al.(2024)Vystr{\v c}ilov{\'a}, Sridhar, Burg,
  Gollisch, and Ecker]{Vystrcilova2024}
Michaela Vystr{\v c}ilov{\'a}, Shashwat Sridhar, Max~F. Burg, Tim Gollisch, and
  Alexander~S. Ecker.
\newblock Convolutional neural network models of the primate retina reveal
  adaptation to natural stimulus statistics.
\newblock \emph{bioRxiv}, 2024.
\newblock \doi{10.1101/2024.03.06.583740}.
\newblock URL
  \url{https://www.biorxiv.org/content/early/2024/03/09/2024.03.06.583740}.

\bibitem[Walker et~al.(2019)Walker, Sinz, Cobos, Muhammad, Froudarakis, Fahey,
  Ecker, Reimer, Pitkow, and Tolias]{walker2019inception}
Edgar~Y Walker, Fabian~H Sinz, Erick Cobos, Taliah Muhammad, Emmanouil
  Froudarakis, Paul~G Fahey, Alexander~S Ecker, Jacob Reimer, Xaq Pitkow, and
  Andreas~S Tolias.
\newblock Inception loops discover what excites neurons most using deep
  predictive models.
\newblock \emph{Nature neuroscience}, 22\penalty0 (12):\penalty0 2060--2065,
  2019.

\bibitem[Franke et~al.(2022)Franke, Willeke, Ponder, Galdamez, Zhou, Muhammad,
  Patel, Froudarakis, Reimer, Sinz, et~al.]{franke2022state}
Katrin Franke, Konstantin~F Willeke, Kayla Ponder, Mario Galdamez, Na~Zhou,
  Taliah Muhammad, Saumil Patel, Emmanouil Froudarakis, Jacob Reimer, Fabian~H
  Sinz, et~al.
\newblock State-dependent pupil dilation rapidly shifts visual feature
  selectivity.
\newblock \emph{Nature}, 610\penalty0 (7930):\penalty0 128--134, 2022.

\bibitem[H{\"o}fling et~al.(2024)H{\"o}fling, Szatko, Behrens, Deng, Qiu,
  Klindt, Jessen, Schwartz, Bethge, Berens, et~al.]{hofling2024chromatic}
Larissa H{\"o}fling, Klaudia~P Szatko, Christian Behrens, Yuyao Deng, Yongrong
  Qiu, David~Alexander Klindt, Zachary Jessen, Gregory~W Schwartz, Matthias
  Bethge, Philipp Berens, et~al.
\newblock A chromatic feature detector in the retina signals visual context
  changes.
\newblock \emph{Elife}, 13:\penalty0 e86860, 2024.

\bibitem[Tong et~al.(2023)Tong, da~Silva, Lin, Ghosh, Wilsenach, Cianfarano,
  Bashivan, Richards, and Trenholm]{Tong2023.11.03.565500}
Rudi Tong, Ronan da~Silva, Dongyan Lin, Arna Ghosh, James Wilsenach, Erica
  Cianfarano, Pouya Bashivan, Blake Richards, and Stuart Trenholm.
\newblock The feature landscape of visual cortex.
\newblock \emph{bioRxiv}, 2023.
\newblock \doi{10.1101/2023.11.03.565500}.
\newblock URL
  \url{https://www.biorxiv.org/content/early/2023/11/05/2023.11.03.565500}.

\bibitem[Ustyuzhaninov et~al.(2022)Ustyuzhaninov, Burg, Cadena, Fu, Muhammad,
  Ponder, Froudarakis, Ding, Bethge, Tolias, et~al.]{ivan}
Ivan Ustyuzhaninov, Max~F Burg, Santiago~A Cadena, Jiakun Fu, Taliah Muhammad,
  Kayla Ponder, Emmanouil Froudarakis, Zhiwei Ding, Matthias Bethge, Andreas~S
  Tolias, et~al.
\newblock Digital twin reveals combinatorial code of non-linear computations in
  the mouse primary visual cortex.
\newblock \emph{bioRxiv}, pages 2022--02, 2022.

\bibitem[Burg et~al.(2024)Burg, Zenkel, Vystrčilová, Oesterle, Höfling,
  Willeke, Lause, Müller, Fahey, Ding, Restivo, Sridhar, Gollisch, Berens,
  Tolias, Euler, Bethge, and
  Ecker]{burg2024discriminativestimulifunctionalcell}
Max~F. Burg, Thomas Zenkel, Michaela Vystrčilová, Jonathan Oesterle, Larissa
  Höfling, Konstantin~F. Willeke, Jan Lause, Sarah Müller, Paul~G. Fahey,
  Zhiwei Ding, Kelli Restivo, Shashwat Sridhar, Tim Gollisch, Philipp Berens,
  Andreas~S. Tolias, Thomas Euler, Matthias Bethge, and Alexander~S. Ecker.
\newblock Most discriminative stimuli for functional cell type clustering,
  2024.
\newblock URL \url{https://arxiv.org/abs/2401.05342}.

\bibitem[Hubert and Arabie(1985)]{ARI}
Lawrence Hubert and Phipps Arabie.
\newblock Comparing partitions.
\newblock \emph{Journal of classification}, 2:\penalty0 193--218, 1985.

\bibitem[Xie et~al.(2016)Xie, Girshick, and Farhadi]{DEC}
Junyuan Xie, Ross Girshick, and Ali Farhadi.
\newblock Unsupervised deep embedding for clustering analysis, 2016.
\newblock URL \url{https://arxiv.org/abs/1511.06335}.

\bibitem[McLachlan and Peel(2000{\natexlab{a}})]{tMM}
Geoffrey~J. McLachlan and David Peel.
\newblock Robust mixture modelling using the t distribution.
\newblock \emph{Statistical Science}, 15\penalty0 (1):\penalty0 1--19,
  2000{\natexlab{a}}.
\newblock URL \url{https://people.smp.uq.edu.au/GeoffMcLachlan/pm_sc00.pdf}.

\bibitem[Yamins et~al.(2014)Yamins, Hong, Cadieu, Solomon, Seibert, and
  DiCarlo]{yamins2014performance}
Daniel~LK Yamins, Ha~Hong, Charles~F Cadieu, Ethan~A Solomon, Darren Seibert,
  and James~J DiCarlo.
\newblock Performance-optimized hierarchical models predict neural responses in
  higher visual cortex.
\newblock \emph{Proceedings of the national academy of sciences}, 111\penalty0
  (23):\penalty0 8619--8624, 2014.

\bibitem[Cadieu et~al.(2014)Cadieu, Hong, Yamins, Pinto, Ardila, Solomon,
  Majaj, and DiCarlo]{cadieu2014deep}
Charles~F Cadieu, Ha~Hong, Daniel~LK Yamins, Nicolas Pinto, Diego Ardila,
  Ethan~A Solomon, Najib~J Majaj, and James~J DiCarlo.
\newblock Deep neural networks rival the representation of primate it cortex
  for core visual object recognition.
\newblock \emph{PLoS computational biology}, 10\penalty0 (12):\penalty0
  e1003963, 2014.

\bibitem[Cadena et~al.(2019)Cadena, Denfield, Walker, Gatys, Tolias, Bethge,
  and Ecker]{cadena2019deep}
Santiago~A Cadena, George~H Denfield, Edgar~Y Walker, Leon~A Gatys, Andreas~S
  Tolias, Matthias Bethge, and Alexander~S Ecker.
\newblock Deep convolutional models improve predictions of macaque v1 responses
  to natural images.
\newblock \emph{PLoS computational biology}, 15\penalty0 (4):\penalty0
  e1006897, 2019.

\bibitem[Pogoncheff et~al.(2023)Pogoncheff, Granley, and
  Beyeler]{pogoncheff2023explaining}
Galen Pogoncheff, Jacob Granley, and Michael Beyeler.
\newblock Explaining v1 properties with a biologically constrained deep
  learning architecture.
\newblock \emph{Advances in Neural Information Processing Systems},
  36:\penalty0 13908--13930, 2023.

\bibitem[Batty et~al.(2016)Batty, Merel, Brackbill, Heitman, Sher, Litke,
  Chichilnisky, and Paninski]{Batty2016MultilayerRN}
Eleanor Batty, Josh Merel, Nora Brackbill, Alexander Heitman, Alexander Sher,
  Alan~M. Litke, E.~J. Chichilnisky, and Liam Paninski.
\newblock Multilayer recurrent network models of primate retinal ganglion cell
  responses.
\newblock In \emph{International Conference on Learning Representations}, 2016.
\newblock URL \url{https://api.semanticscholar.org/CorpusID:39002941}.

\bibitem[McIntosh et~al.(2016)McIntosh, Maheswaranathan, Nayebi, Ganguli, and
  Baccus]{mcintosh2016deep}
Lane~T. McIntosh, Niru Maheswaranathan, Aran Nayebi, Surya Ganguli, and
  Stephen~A. Baccus.
\newblock Deep learning models of the retinal response to natural scenes.
\newblock In \emph{Advances in Neural Information Processing Systems},
  volume~29, pages 1369--1377. Curran Associates, Inc., 2016.

\bibitem[Sinz et~al.(2018)Sinz, Ecker, Fahey, Walker, Cobos, Froudarakis,
  Yatsenko, Pitkow, Reimer, and Tolias]{sinz2018stimulus}
Fabian Sinz, Alexander~S Ecker, Paul Fahey, Edgar Walker, Erick Cobos,
  Emmanouil Froudarakis, Dimitri Yatsenko, Zachary Pitkow, Jacob Reimer, and
  Andreas Tolias.
\newblock Stimulus domain transfer in recurrent models for large scale cortical
  population prediction on video.
\newblock \emph{Advances in neural information processing systems}, 31, 2018.

\bibitem[Bashiri et~al.(2021)Bashiri, Walker, Lurz, Jagadish, Muhammad, Ding,
  Ding, Tolias, and Sinz]{NEURIPS2021_84a529a9}
Mohammad Bashiri, Edgar Walker, Konstantin-Klemens Lurz, Akshay Jagadish,
  Taliah Muhammad, Zhiwei Ding, Zhuokun Ding, Andreas Tolias, and Fabian Sinz.
\newblock A flow-based latent state generative model of neural population
  responses to natural images.
\newblock In M.~Ranzato, A.~Beygelzimer, Y.~Dauphin, P.S. Liang, and J.~Wortman
  Vaughan, editors, \emph{Advances in Neural Information Processing Systems},
  volume~34, pages 15801--15815. Curran Associates, Inc., 2021.
\newblock URL
  \url{https://proceedings.neurips.cc/paper_files/paper/2021/file/84a529a92de322be42dd3365afd54f91-Paper.pdf}.

\bibitem[Tan et~al.(2011)Tan, Brown, Scholl, Mohanty, and Priebe]{Tan2011}
Andrew~Y. Tan, Brian~D. Brown, Benjamin Scholl, Debarghya Mohanty, and
  Nicholas~J. Priebe.
\newblock Orientation selectivity of synaptic input to neurons in mouse and cat
  primary visual cortex.
\newblock \emph{Journal of Neuroscience}, 31\penalty0 (34):\penalty0
  12339--12350, 2011.
\newblock \doi{10.1523/JNEUROSCI.2039-11.2011}.
\newblock Erratum in: J Neurosci. 2011 Oct 12;31(41):14832.

\bibitem[Li et~al.(2023)Li, Cornacchia, Rochefort, and Onken]{li2023v1t}
Bryan~M Li, Isabel~M Cornacchia, Nathalie~L Rochefort, and Arno Onken.
\newblock V1t: large-scale mouse v1 response prediction using a vision
  transformer.
\newblock \emph{arXiv preprint arXiv:2302.03023}, 2023.

\bibitem[Klindt et~al.(2017)Klindt, Ecker, Euler, and Bethge]{whatandwhere}
David Klindt, Alexander~S Ecker, Thomas Euler, and Matthias Bethge.
\newblock Neural system identification for large populations separating
  “what” and “where”.
\newblock \emph{Advances in neural information processing systems}, 30, 2017.

\bibitem[Lurz et~al.(2020)Lurz, Bashiri, Willeke, Jagadish, Wang, Walker,
  Cadena, Muhammad, Cobos, Tolias, Ecker, and Sinz]{Lurz}
Konstantin-Klemens Lurz, Mohammad Bashiri, Konstantin Willeke, Akshay~K.
  Jagadish, Eric Wang, Edgar~Y. Walker, Santiago~A. Cadena, Taliah Muhammad,
  Erick Cobos, Andreas~S. Tolias, Alexander~S. Ecker, and Fabian~H. Sinz.
\newblock Generalization in data-driven models of primary visual cortex.
\newblock \emph{bioRxiv}, 2020.
\newblock \doi{10.1101/2020.10.05.326256}.
\newblock URL
  \url{https://www.biorxiv.org/content/early/2020/10/07/2020.10.05.326256}.

\bibitem[Pierzchlewicz et~al.(2023)Pierzchlewicz, Willeke, Nix, Elumalai,
  Restivo, Shinn, Nealley, Rodriguez, Patel, Franke,
  et~al.]{pierzchlewicz2023energy}
Pawel Pierzchlewicz, Konstantin Willeke, Arne Nix, Pavithra Elumalai, Kelli
  Restivo, Tori Shinn, Cate Nealley, Gabrielle Rodriguez, Saumil Patel, Katrin
  Franke, et~al.
\newblock Energy guided diffusion for generating neurally exciting images.
\newblock \emph{Advances in Neural Information Processing Systems},
  36:\penalty0 32574--32601, 2023.

\bibitem[Ustyuzhaninov et~al.(2019)Ustyuzhaninov, Cadena, Froudarakis, Fahey,
  Walker, Cobos, Reimer, Sinz, Tolias, Bethge,
  et~al.]{ustyuzhaninov2019rotation}
Ivan Ustyuzhaninov, Santiago~A Cadena, Emmanouil Froudarakis, Paul~G Fahey,
  Edgar~Y Walker, Erick Cobos, Jacob Reimer, Fabian~H Sinz, Andreas~S Tolias,
  Matthias Bethge, et~al.
\newblock Rotation-invariant clustering of neuronal responses in primary visual
  cortex.
\newblock In \emph{International Conference on Learning Representations}, 2019.

\bibitem[MacQueen(1967)]{kmeans}
James MacQueen.
\newblock Some methods for classification and analysis of multivariate
  observations.
\newblock In \emph{Proceedings of the Fifth Berkeley Symposium on Mathematical
  Statistics and Probability}, pages 281--297, 1967.

\bibitem[McLachlan and Peel(2000{\natexlab{b}})]{mclachlan2000finite}
Geoffrey~J McLachlan and David Peel.
\newblock \emph{Finite mixture models}.
\newblock John Wiley \& Sons, 2000{\natexlab{b}}.

\bibitem[Dayan and Abbott(2005)]{Poisson}
Peter Dayan and Laurence~F. Abbott.
\newblock \emph{Theoretical Neuroscience: Computational and Mathematical
  Modeling of Neural Systems}.
\newblock MIT Press, 2005.

\bibitem[Vintch et~al.(2015)Vintch, Movshon, and Simoncelli]{corr1}
Brett Vintch, J~Anthony Movshon, and Eero~P Simoncelli.
\newblock A convolutional subunit model for neuronal responses in macaque v1.
\newblock \emph{Journal of Neuroscience}, 35\penalty0 (44):\penalty0
  14829--14841, 2015.

\bibitem[Burg et~al.(2021)Burg, Cadena, Denfield, Walker, Tolias, Bethge, and
  Ecker]{burg2021learning}
Max~F Burg, Santiago~A Cadena, George~H Denfield, Edgar~Y Walker, Andreas~S
  Tolias, Matthias Bethge, and Alexander~S Ecker.
\newblock Learning divisive normalization in primary visual cortex.
\newblock \emph{PLoS computational biology}, 17\penalty0 (6):\penalty0
  e1009028, 2021.

\bibitem[Maaten and Hinton(2008)]{tSNE}
Laurens van~der Maaten and Geoffrey Hinton.
\newblock Visualizing data using t-sne.
\newblock \emph{Journal of machine learning research}, 9\penalty0
  (Nov):\penalty0 2579--2605, 2008.

\bibitem[Linderman and Steinerberger(2017)]{tsne_setting}
George~C. Linderman and Stefan Steinerberger.
\newblock Clustering with t-sne, provably, 2017.
\newblock URL \url{https://arxiv.org/abs/1706.02582}.

\bibitem[Masri et~al.(2019)Masri, Percival, Koizumi, Martin, and
  Gr{\"u}nert]{masri2019}
Rania~A Masri, Kumiko~A Percival, Amane Koizumi, Paul~R Martin, and Ulrike
  Gr{\"u}nert.
\newblock Survey of retinal ganglion cell morphology in marmoset.
\newblock \emph{Journal of Comparative Neurology}, 527\penalty0 (1):\penalty0
  236--258, 2019.

\bibitem[Sridhar and Gollisch(2025)]{sridhar_dataset_2025}
Shashwat Sridhar and Tim Gollisch.
\newblock Dataset - {Marmoset} retinal ganglion cell responses to naturalistic
  movies and spatiotemporal white noise, April 2025.
\newblock URL \url{https://doi.gin.g-node.org/10.12751/g-node.3dfiti}.

\bibitem[Zhao et~al.(2020)Zhao, Klindt, Maia~Chagas, Szatko, Rogerson, Protti,
  Behrens, Dalkara, Schubert, Bethge, et~al.]{zhao2020temporal}
Zhijian Zhao, David~A Klindt, Andr{\'e} Maia~Chagas, Klaudia~P Szatko, Luke
  Rogerson, Dario~A Protti, Christian Behrens, Deniz Dalkara, Timm Schubert,
  Matthias Bethge, et~al.
\newblock The temporal structure of the inner retina at a single glance.
\newblock \emph{Scientific reports}, 10\penalty0 (1):\penalty0 4399, 2020.

\bibitem[Gouwens et~al.(2019)Gouwens, Sorensen, Berg, Lee, Jarsky, Ting,
  Sunkin, Feng, Anastassiou, Barkan, et~al.]{gouwens2019classification}
Nathan~W Gouwens, Staci~A Sorensen, Jim Berg, Changkyu Lee, Tim Jarsky,
  Jonathan Ting, Susan~M Sunkin, David Feng, Costas~A Anastassiou, Eliza
  Barkan, et~al.
\newblock Classification of electrophysiological and morphological neuron types
  in the mouse visual cortex.
\newblock \emph{Nature neuroscience}, 22\penalty0 (7):\penalty0 1182--1195,
  2019.

\bibitem[Willeke et~al.(2023)Willeke, Restivo, Franke, Nix, Cadena, Shinn,
  et~al.]{willeke2023deep}
KF~Willeke, K~Restivo, K~Franke, AF~Nix, SA~Cadena, T~Shinn, et~al.
\newblock Deep learning-driven characterization of single cell tuning in
  primate visual area v4 unveils topological organization. biorxiv. 2023.
\newblock \emph{doi. org/10.1101/2023.05}, 12, 2023.

\bibitem[Weiler et~al.(2023)Weiler, Guggiana~Nilo, Bonhoeffer, H{\"u}bener,
  Rose, and Scheuss]{Weiler2023}
Sebastian Weiler, Daniel Guggiana~Nilo, Tobias Bonhoeffer, Mark H{\"u}bener,
  Tobias Rose, and Volker Scheuss.
\newblock Functional and structural features of l2/3 pyramidal cells
  continuously covary with pial depth in mouse visual cortex.
\newblock \emph{Cerebral Cortex}, 33\penalty0 (7):\penalty0 3715--3733, 2023.
\newblock \doi{10.1093/cercor/bhac303}.

\bibitem[Sridhar et~al.(2025)Sridhar, Vystr{\v c}ilov{\'a}, Khani, Karamanlis,
  Schreyer, Ramakrishna, Kr{\"u}ppel, Zapp, Mietsch, Ecker, and
  Gollisch]{Sridhar2025}
Shashwat Sridhar, Michaela Vystr{\v c}ilov{\'a}, Mohammad~H. Khani, Dimokratis
  Karamanlis, Helene~M. Schreyer, Varsha Ramakrishna, Steffen Kr{\"u}ppel,
  S{\"o}ren~J. Zapp, Matthias Mietsch, Alexander~S. Ecker, and Tim Gollisch.
\newblock Modeling spatial contrast sensitivity in responses of primate retinal
  ganglion cells to natural movies.
\newblock \emph{bioRxiv}, 2025.
\newblock \doi{10.1101/2024.03.05.583449}.
\newblock URL
  \url{https://www.biorxiv.org/content/early/2025/04/09/2024.03.05.583449}.

\bibitem[Fowlkes and Mallows(1983)]{fowlkes1983method}
Edward~B Fowlkes and Colin~L Mallows.
\newblock A method for comparing two hierarchical clusterings.
\newblock \emph{Journal of the American statistical association}, 78\penalty0
  (383):\penalty0 553--569, 1983.

\bibitem[Rosenberg and Hirschberg(2007)]{rosenberg2007v}
Andrew Rosenberg and Julia Hirschberg.
\newblock V-measure: A conditional entropy-based external cluster evaluation
  measure.
\newblock In \emph{Proceedings of the 2007 joint conference on empirical
  methods in natural language processing and computational natural language
  learning (EMNLP-CoNLL)}, pages 410--420, 2007.

\bibitem[Strehl and Ghosh(2002)]{strehl2002cluster}
Alexander Strehl and Joydeep Ghosh.
\newblock Cluster ensembles---a knowledge reuse framework for combining
  multiple partitions.
\newblock \emph{Journal of machine learning research}, 3\penalty0
  (Dec):\penalty0 583--617, 2002.

\end{thebibliography}
\newpage
\appendix

\section{Acknowledgments}
We thank Suhas Shrinivasan, Max F. Burg, Larissa Höfling, Thomas Zenkel, Konstantin F Willeke, Fabian Sinz.

This work was supported by the Deutsche Forschungsgemeinschaft (DFG, German Research Foundation) -- project IDs 432680300 (SFB 1456, project B05) and 515774656 -- and by the European Research Council (ERC) under the European Union’s Horizon 2020 research and innovation programme (grant agreement number 101041669).
Views and opinions expressed are however those of the authors only and do not necessarily reflect those of the European Union or the European Research Council Executive Agency. 
The work was supported by the Intelligence Advanced Research Projects Activity (IARPA) via Department of Interior/ Interior Business Center (DoI/IBC) contract number D16PC00003. 
The U.S. Government is authorized to reproduce and distribute reprints for Governmental purposes notwithstanding any copyright annotation thereon. 
AST acknowledges support from National Institute of Mental Health and National Institute of Neurological Disorders And Stroke under Award Number U19MH114830 and National Eye Institute award numbers R01 EY026927 and Core Grant for Vision Research T32-EY-002520-37. 
Disclaimer: The views and conclusions contained herein are those of the authors and should not be interpreted as necessarily representing the official policies or endorsements, either expressed or implied, of IARPA, DoI/IBC, or the U.S. Government. 
We gratefully acknowledge the computing time granted by the Resource Allocation Board and provided on the supercomputer Emmy/Grete at NHR-Nord@Göttingen as part of the NHR infrastructure. 
The calculations for this research were conducted with computing resources under the projects nim00010 and nim00012.

\section{Appendix}

\subsection{Retina gagnlion cells}

\label{app:retinaGagnlion}

\begin{wrapfigure}{r}{0.5\textwidth}
    \centering
    \begin{subfigure}[t]{0.22\textwidth}
        \begin{tikzpicture}
            \node[anchor=north west, inner sep=0] (img) at (0,0) 
                {\includegraphics[width=\textwidth]{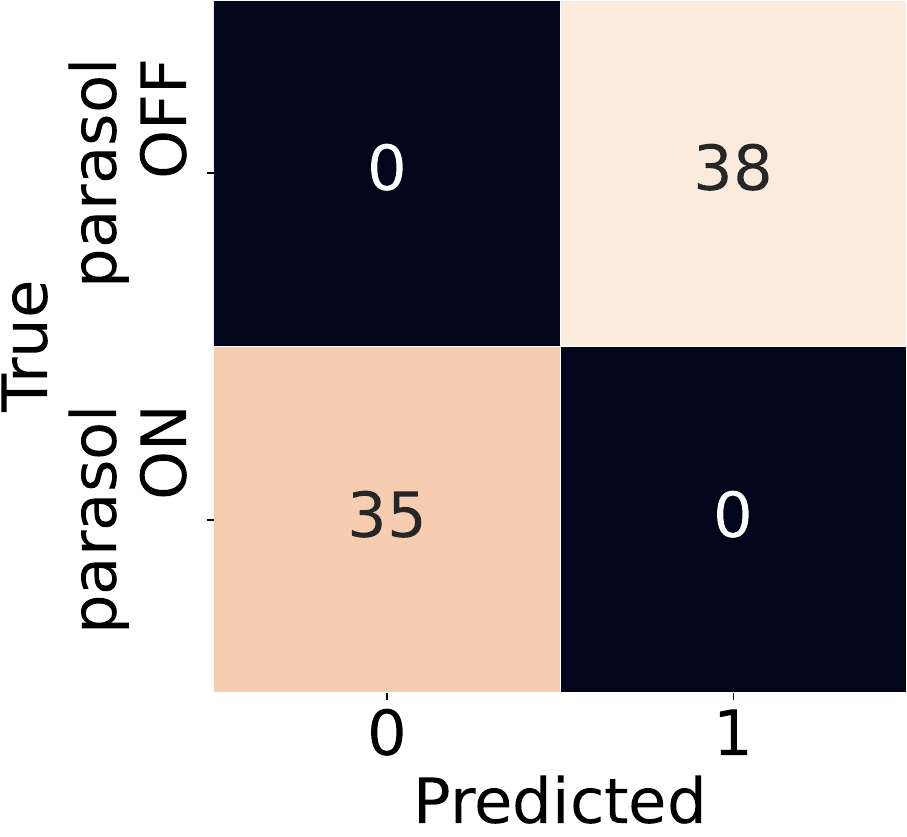}};
            \node[anchor=north west, inner sep=1pt, xshift=0em, yshift=0.8em] at (img.north west)
                {{\fontfamily{qhv}\selectfont\bfseries A}};
        \end{tikzpicture}
    \end{subfigure}%
    \hfill
    \vspace{2mm}
    \begin{subfigure}[t]{0.27\textwidth}
        \begin{tikzpicture}
            \node[anchor=north west, inner sep=0] (img) at (0,0)
                {\includegraphics[width=\textwidth]{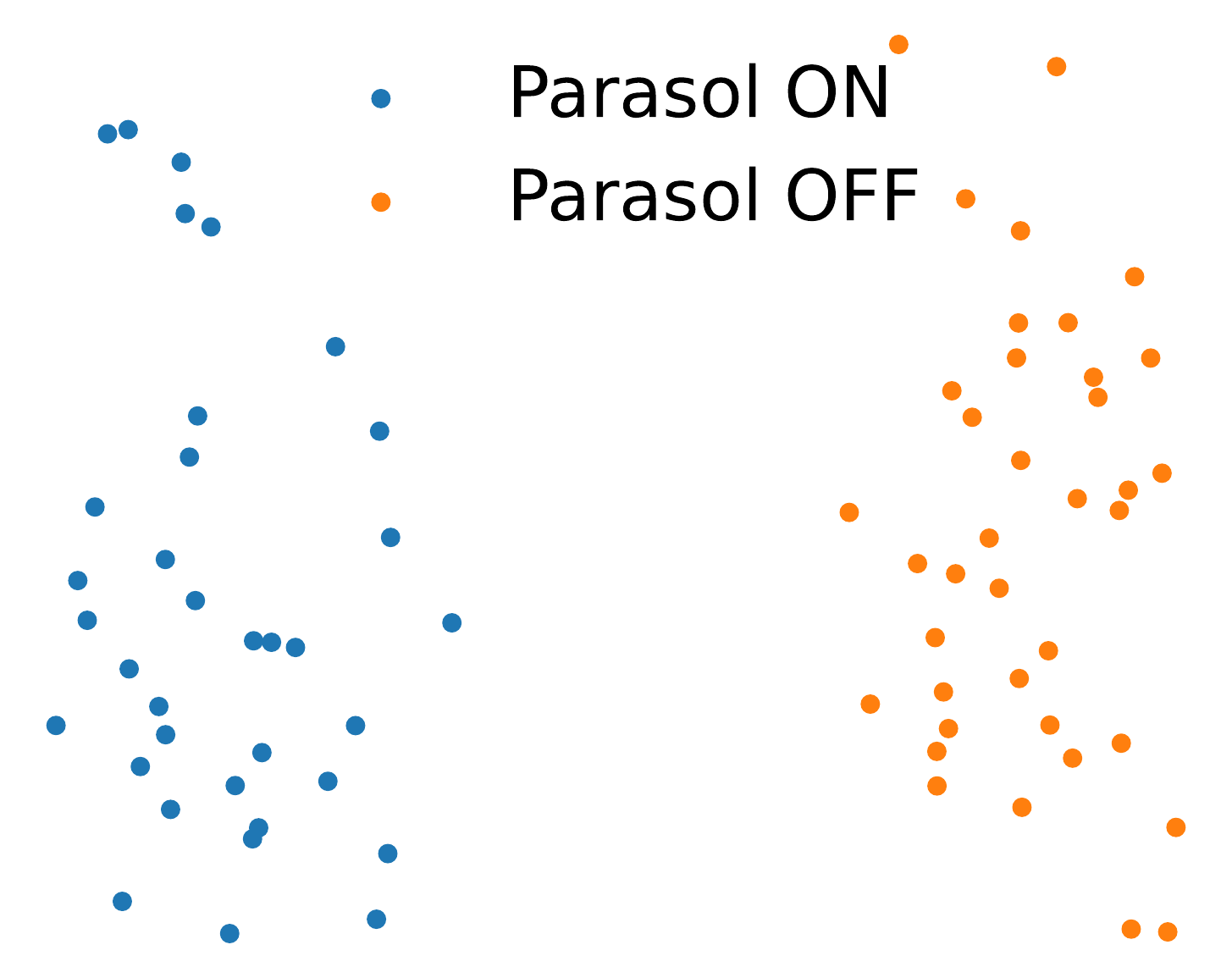}};
            \node[anchor=north west, inner sep=1pt, xshift=0.8em, yshift=0.5em] at (img.north west)
                {{\fontfamily{qhv}\selectfont\bfseries B}};
        \end{tikzpicture}
    \end{subfigure}%
        \caption{
    \textbf{A}: DECEMber predictions. Pretraining length: 20 epochs. Same predictions for GMM and k-means. All methods have ARI 1.
    \textbf{B}: $t$-SNE projections of the corresponding cells.}
    \label{fig:retina_marmoset4}
\end{wrapfigure}

To select reliable cells from the marmoset RGCs dataset \cite{sridhar_dataset_2025}, we used the same reliability assessment of each cell's responses to visual stimuli as in \cite{Sridhar2025}; only reliable cells were used for model training. The model architecture was also taken from \cite{Vystrcilova2024}. For clustering evaluations using DECEMber, k-means, and GMM, we considered only cells for which cell-type labels were available.

The dataset contains recordings from two different retinas of male marmosets. The second retina (not analyzed in the main part of this paper) includes 38 parasol-OFF and 35 parasol-ON cells which are well separable (\Cref{fig:retina_marmoset4}B). We trained our models on all reliable cells from this retina as well and tested DECEMber with varying pretraining lengths (which did not affect cluster consistency). All three clustering methods—GMM, k-means, and DECEMber—successfully and robustly identified the two cell types, as visualized in \cref{fig:retina_marmoset4}A with ARI=1.

\paragraph{Cell type labels}
We used the same cell-type classification procedure as in \cite{Sridhar2025} (Methods section 4.5), clustering the cells using the KMeans++ algorithm on features extracted from receptive-field estimates obtained using spike-triggered averaging from responses to spatiotemporal white-noise, and from autocorrelograms computed on responses to white-noise and naturalistic movies.
The cell-type labels in our analysis differ from the ones used in the original publication because we did not exclude cells that violated the tiling of spatial receptive fields. 

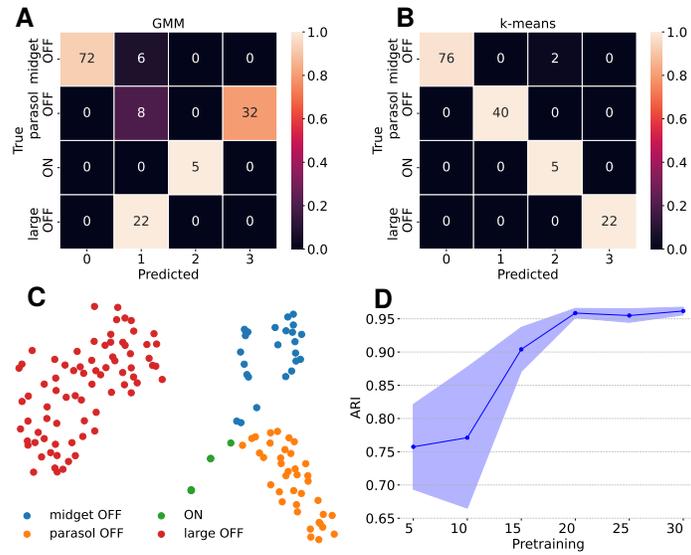
\begin{figure}[h]
    \centering
    \begin{subfigure}[t]{0.3\textwidth}
        \begin{tikzpicture}
            \node[anchor=north west, inner sep=0] (img) at (0,0) 
                {\includegraphics[width=\textwidth]{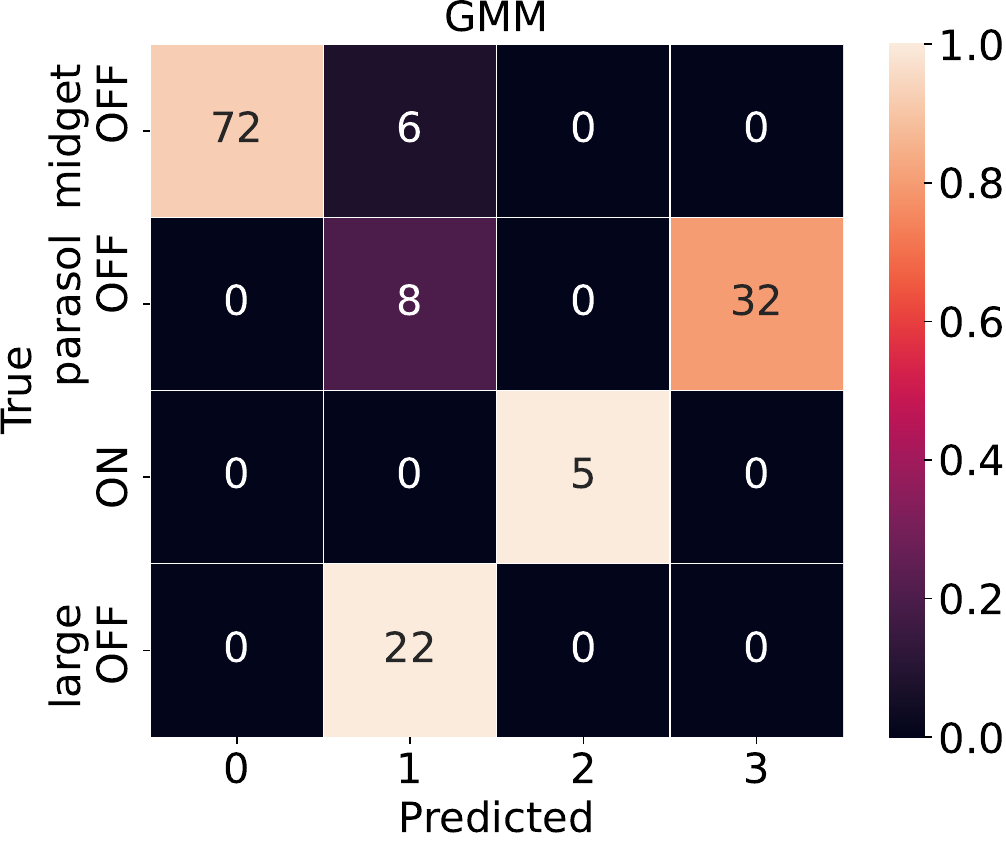}};
            \node[anchor=north west, inner sep=1pt, xshift=0em, yshift=0.5em] at (img.north west)
                {{\fontfamily{qhv}\selectfont\bfseries A}};
        \end{tikzpicture}
    \end{subfigure}%
    \hspace{5mm}
    \begin{subfigure}[t]{0.3\textwidth}
        \begin{tikzpicture}
            \node[anchor=north west, inner sep=0] (img) at (0,0)
                {\includegraphics[width=\textwidth]{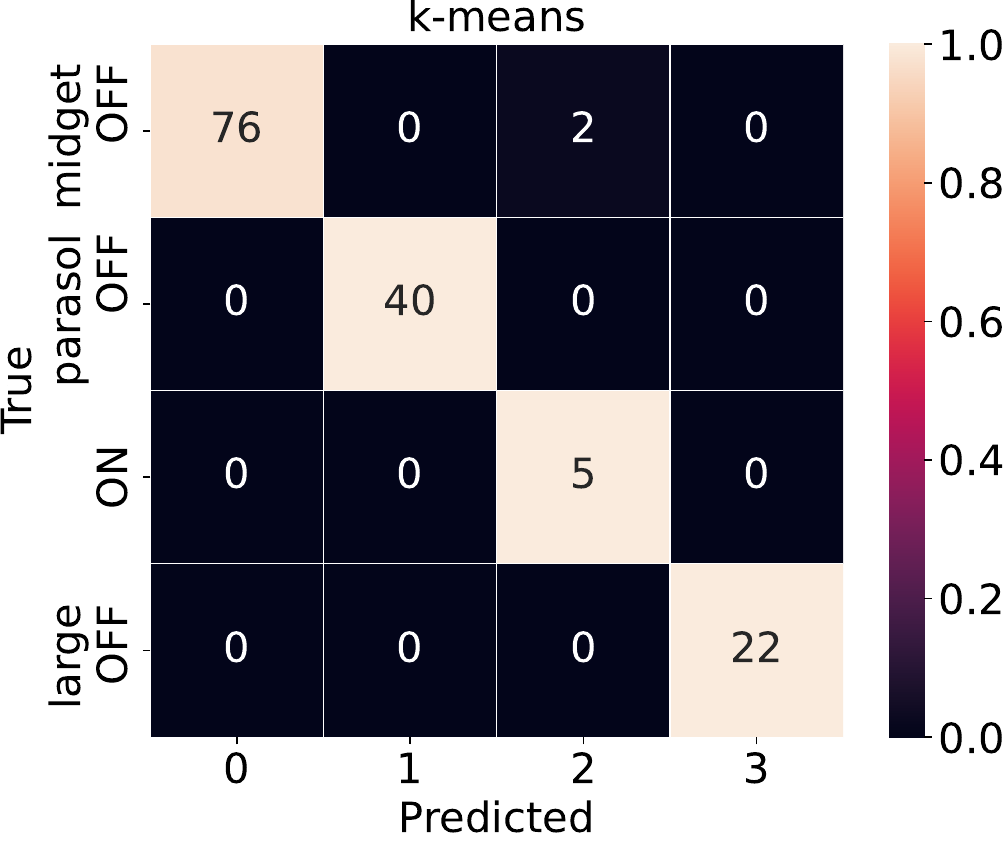}};
            \node[anchor=north west, inner sep=1pt, xshift=0.8em, yshift=0.5em] at (img.north west)
                {{\fontfamily{qhv}\selectfont\bfseries B}};
        \end{tikzpicture}
    \end{subfigure}%
    \hspace{5mm}
        \begin{subfigure}[t]{0.33\textwidth}
        \begin{tikzpicture}
            \node[anchor=north west, inner sep=0] (img) at (0,0)
                {\includegraphics[width=\textwidth]{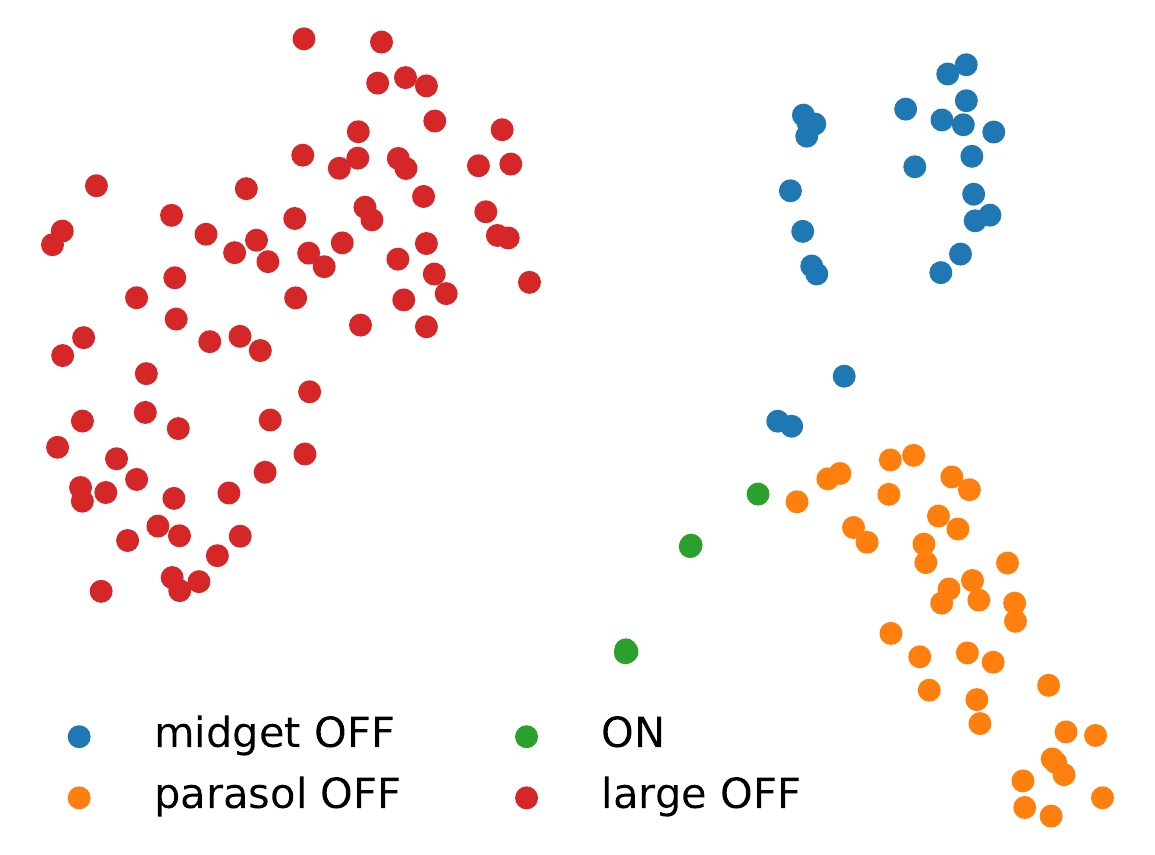}};
            \node[anchor=north west, inner sep=1pt, xshift=0.8em, yshift=0.4em] at (img.north west)
                {{\fontfamily{qhv}\selectfont\bfseries C}};
        \end{tikzpicture}
    \end{subfigure}%
    \begin{subfigure}[t]{0.33\textwidth}
        \begin{tikzpicture}
            \node[anchor=north west, inner sep=0] (img) at (0,0)
                {\includegraphics[width=\textwidth]{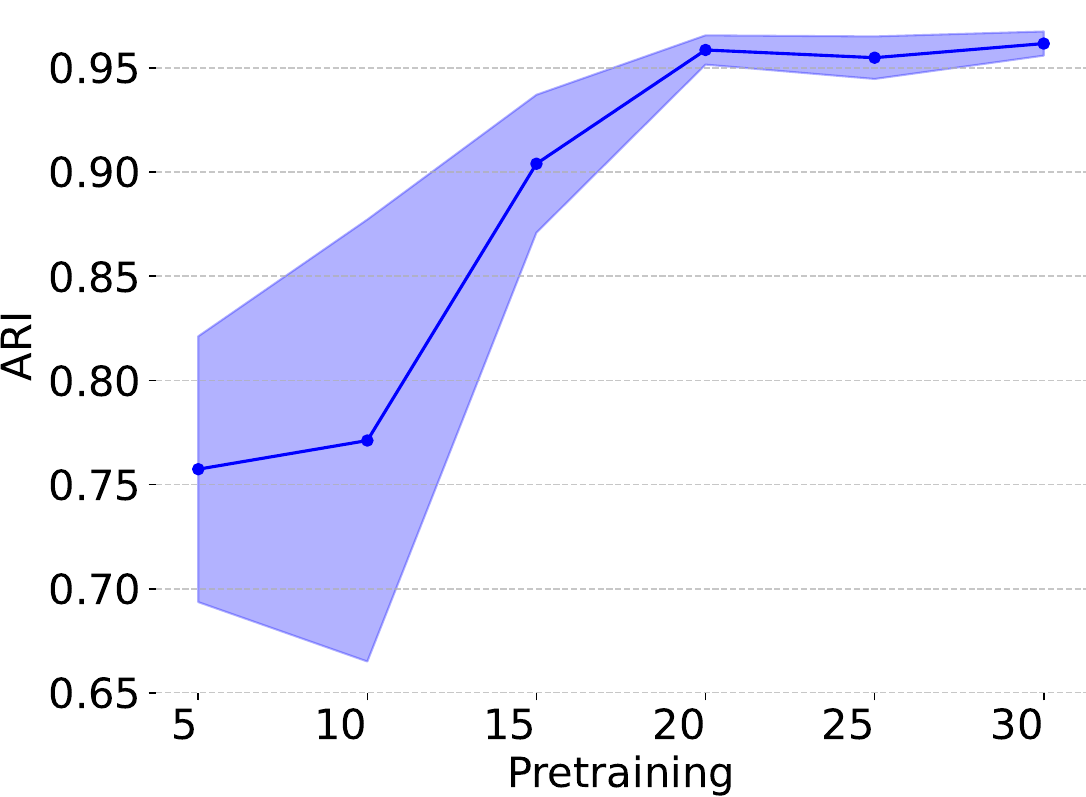}};
            \node[anchor=north west, inner sep=1pt, xshift=0.8em, yshift=0.4em] at (img.north west)
                {{\fontfamily{qhv}\selectfont\bfseries D}};
        \end{tikzpicture}
    \end{subfigure}%
    \caption{
    All plots show evaluations of seed 4 of the trained marmoset RGC model \cite{Vystrcilova2024} for retina 1 used in the main part of this paper.
    \textbf{A}: GMM predictions. 
    \textbf{B}: k-means.   
    \textbf{C}:
    $t$-SNE projections of the corresponding cells. \textbf{D}: ARI for different length of pretraining. Longer pretraining seems to be beneficial with ARI stabilizing after pretraining of 20 epochs.}
    \label{fig:retina_marmoset_additional}
\end{figure}

\subsection{ARI-stability for k-means and GMM on marmoset RGC}


\begin{figure}[H]
\centering
\begin{minipage}{0.33\textwidth}
    \includegraphics[width=\linewidth]{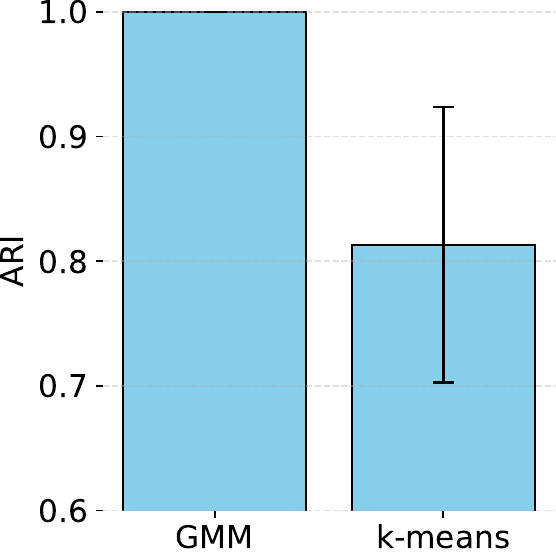}
\end{minipage}
\hspace{1em} 
\begin{minipage}{0.52\textwidth}
    \captionof{figure}{\textbf{Clustering stability of k-means and GMM on retina 1 for marmoset RGC.}
    We started with a single baseline RGC model of retina 1 (seed 2) and performed k-means and GMM clustering (4 clusters each), varying the random seed (42, 10, 100) for both algorithms. Clustering was done on all cells the model was trained on, but ARI was calculated using only labeled cells.}
    \label{fig:seed_dependence}
\end{minipage}
\end{figure}

\subsection{ARI stability for GMM on mouse V1}
\begin{figure}[H]
\centering
\begin{minipage}{0.4\textwidth}
    \includegraphics[width=\linewidth]{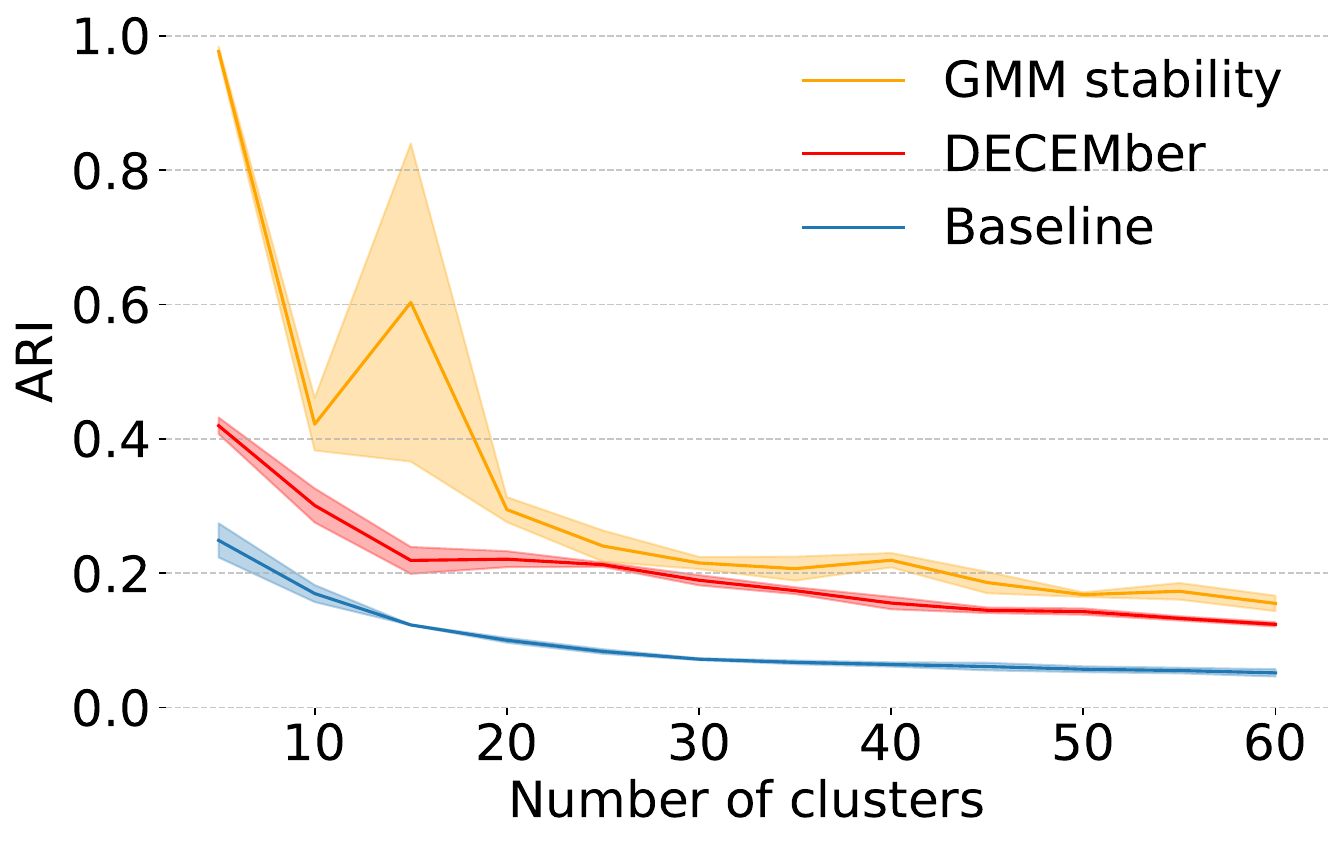}
\end{minipage}
\hspace{1em} 
\begin{minipage}{0.52\textwidth}
    \captionof{figure}{\textbf{Clustering stability of GMM on mouse V1.}
    We trained a baseline model of mouse V1 for one seed (seed=42). 
    We then did GMM for clusters ranging from 5 to 60 with a step size of 5 as the ground truth is not known varying just the seed for the initialization of the GMM but using the embeddings from the same initial model (orange line). 
    We used 3 seeds (42,12,123) for the GMM initialization. 
    The GMM becomes extremely initialization sensitive on exactly same data  with the growing amount of clusters. 
    In comparison we show post-hoc GMM on baseline model trained on 3 different models with seeds (42,10,100) (blue line) and DECEMber with same seeds and again 3 different models, pretraining of 10 epochs, $\beta = 10^5$ (red line). 
    After 25 clusters DECEMber and GMM show almost same stability, while the GMM based on different models but with same GMM initialization seed -- blue line -- is even worse.
    }
    \label{fig:seed_dependence2}
\end{minipage}
\end{figure}

\subsection{Sensorium data details.}
\label{app:sensoriumData}
The model was trained on the SENSORIUM 2022 dataset \cite{sensorium}, which contains neural responses to natural images recorded from seven mice (a total of 54,569 neurons). 
Recordings were made from excitatory neurons in layer 2 and 3 of the primary visual cortex using two-photon calcium imaging. 
In addition to neural activity, the dataset includes five behavioral variables: locomotion speed, pupil size, the instantaneous change in pupil size (estimated via second-order central differences), and horizontal and vertical eye position, all of which are incorporated into the model. 
Three of them -- locomotion speed, pupil size, and the instantaneous change in pupil size -- were appended to the grayscale images and are used as input to the core, while pupil horizontal and vertical position were used as input to the shifter -- a model part shifting the readout receptive field locations depending on where the mouse is looking.
The validation set contains responses to roughly 500 and test set to 5000 images per mouse.

\subsection{Additional clustering metrics show qualitatively consistent results with ARI}

\begin{figure}[h]
    \centering
    \includegraphics[width = \textwidth]{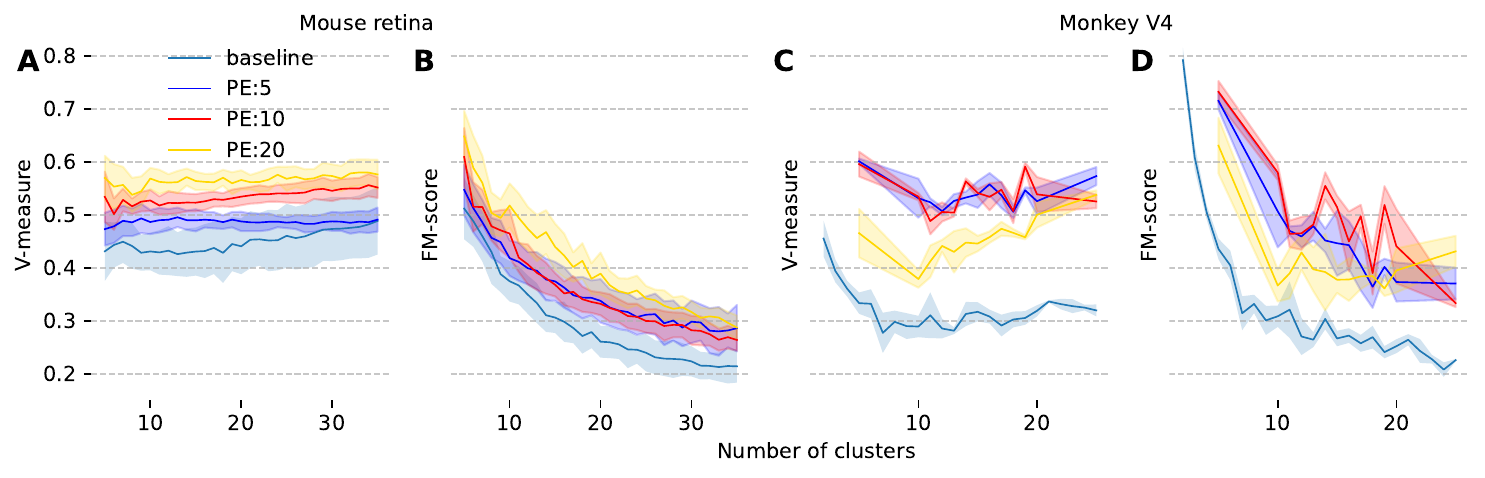}
    \caption{
    Different clustering consistency metrics for monkey V4 and mouse retina datasets. Same as in the main paper, mouse retina is weighted across six models using all neurons, monkey V4 model is trained on a subset of 1000 neurons.
    The order of lines is same as for ARI, confirming its results qualitatively.
    V-measure is biased towards bigger amount of clusters due to the set-based nature.
    }
    \label{fig:fmAdditional}
\end{figure}

\begin{figure}
    \centering
    \includegraphics[width=0.75\linewidth]{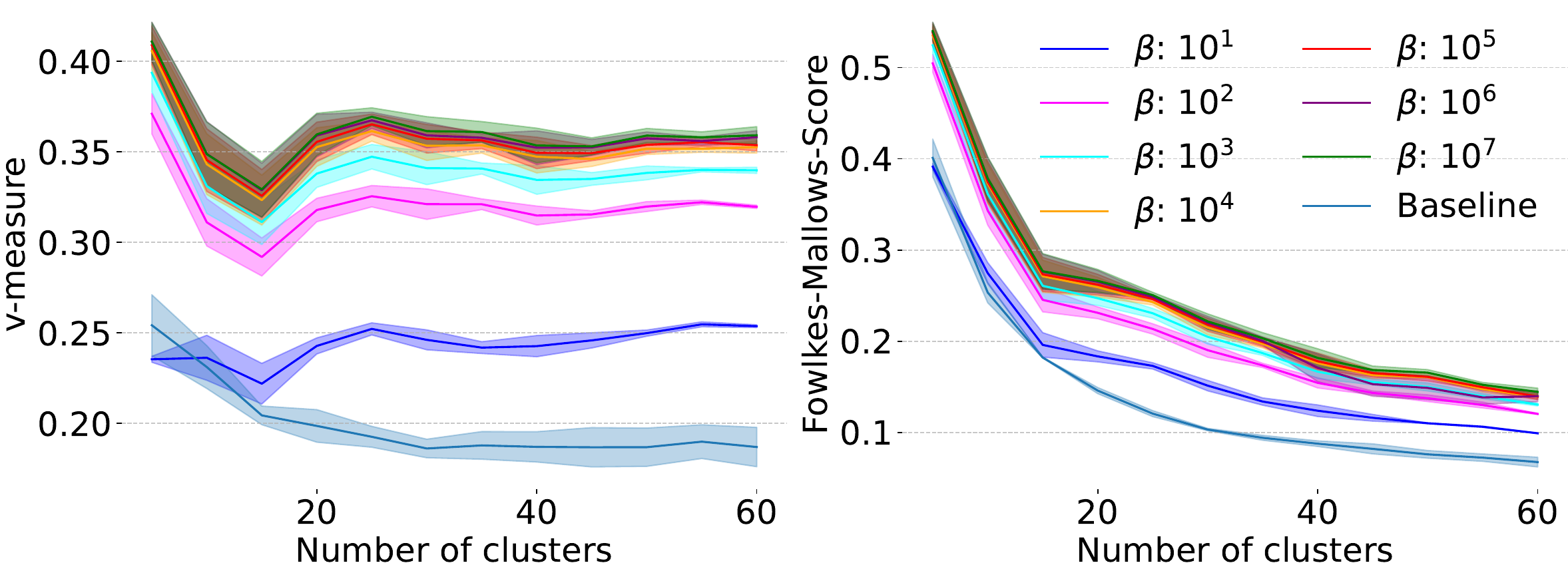}
    \caption{V-measure and Fowlkes-Mallows-score for PE 10, 15 clusters.}
    \label{fig:FM_vscore}
\end{figure}

Clustering quality can be evaluated using metrics beyond ARI. 
ARI measures the similarity between two clusterings by checking whether pairs of points are assigned to the same cluster in both. 
The Fowlkes-Mallows index \cite{fowlkes1983method} (\cref{fig:FM_vscore})  also compares two partitions but does not adjust for chance; it is the geometric mean of precision and recall, based on how consistently point pairs are clustered together.

Other common metrics -- homogeneity, completeness, and V-measure \cite{rosenberg2007v} 
 -- are asymmetric and compare one clustering against a reference (typically ground truth). 
Homogeneity measures whether each cluster contains only members of a single class, while completeness checks whether all members of a given class are assigned to the same cluster. 
Swapping the roles of predicted and true labels interchanges homogeneity and completeness.
V-measure, equivalent to normalized mutual information (NMI \cite{strehl2002cluster}) and it is the harmonic mean of the two.
As in our case we do not have ground truth, we compute the metrics with all possible seed pairs, which leads to homogeneity, completeness, and V-measure being equivalent (\cref{fig:FM_vscore}).

\subsection{Comparison with the rotation equivariant baseline}
\begin{figure}[h]
    \centering
    \includegraphics[width=0.5\textwidth]{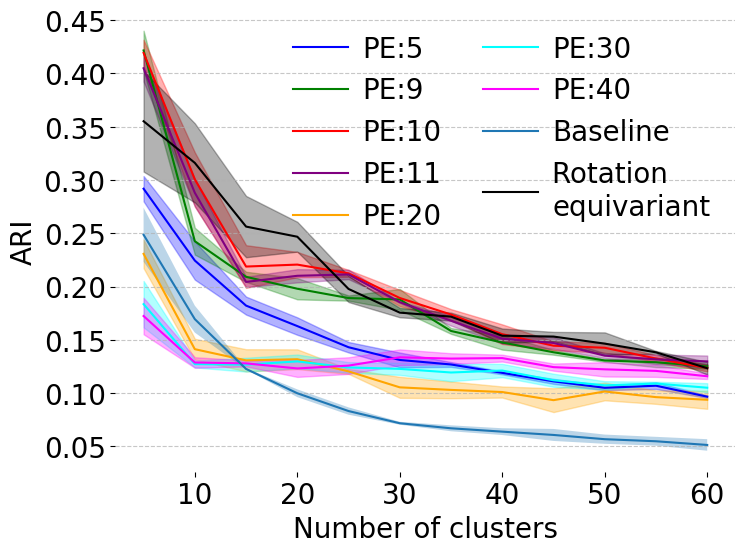}
    \caption{DECEMber, with PE = 10; DECEMber cluster consitency matches the rotation-equivariant model from Turishcheva et al. \cite{turishcheva202}. }
    \label{fig:rotation-equivariant}
\end{figure}

Turishcheva et al. \cite{turishcheva202} is the only work to date that specifically addresses neuronal embedding consistency, and thus serves as our baseline for comparison. 
We use the $\gamma_{\text{lognorm}} = 10$ condition from their paper and compare it to our consistency results in \Cref{fig:rotation-equivariant}. 
Our approach achieves comparable consistency levels while eliminating the need for a rotation-equivariant core, thereby removing the post-hoc alignment step and improving predictive performance from $\approx 38.1\%$  (Fig. 3 A in \cite{turishcheva202}) to $\approx 39.5\%$ (\cref{fig:qualitative}F).


\subsection{Models performances on mouse retina and macaque V4 data}
\label{app:additionalPerformances}

\begin{table}[h!]
\centering
\begin{tabular}{c|c|c|c|c}

 & Baseline & PE 5 & PE 10 & PE 20 \\
\hline
Mouse retina & 0.4727 $\pm$ 0.0008 & 0.4695 $\pm$ 0.0009 & 0.4732 $\pm$ 0.0008 & 0.4727 $\pm$ 0.0009 \\
Macaque V4 & 0.308 $\pm$ 0.004 & 0.308 $\pm$ 0.006 & 0.304 $\pm$ 0.005 & 0.305 $\pm$ 0.003 \\
\hline
\end{tabular}
\caption{Performances on mouse retina and macaque V4 data for the models reported in the main paper (\Cref{sec:generalization}). 
Mouse retina is weighted as described in \Cref{app:lara}.All performances are on the held-out test set. 
The values are averaged across all cluster counts.
Seeds were 42, 101 and 7607. 
For GMM baseline the seed was 42.
}
\label{tab:additionalPerformances}
\end{table}

\subsection{Further analysis of mouse retina data}
\label{app:lara}
\begin{figure}[htb]
\centering
    \includegraphics[width=\linewidth]{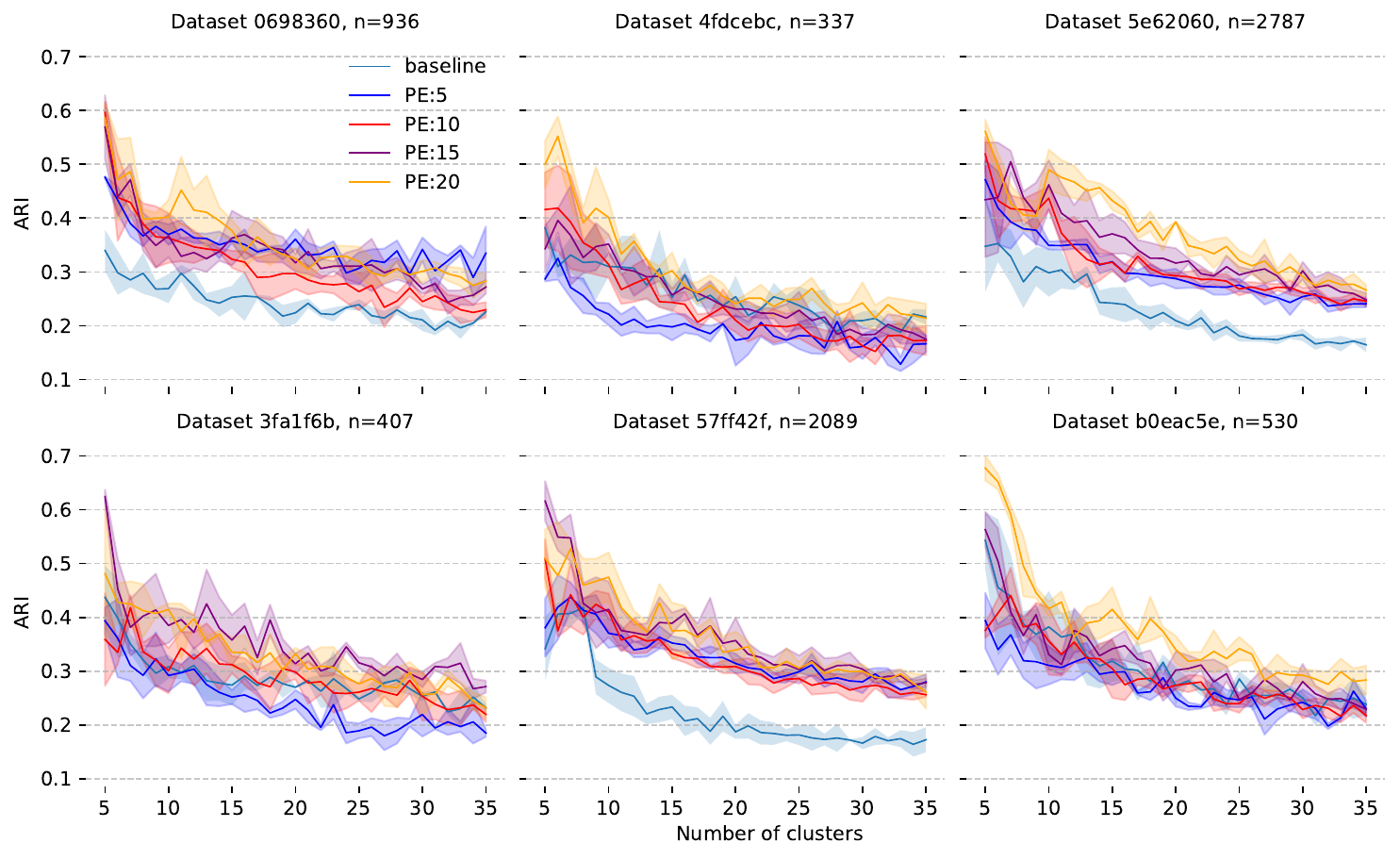}
    \caption{ARI per retina. $n$ is the number of neurons in the model}
    \label{fig:perRetina}
\end{figure}

\paragraph{Averaging across datasets}
For the mean ARI across datasets we weighted ARI lines like $\mu_{\text{total}} = \sum_i w_i \mu_i
$, where $w_i= n_{\text{cur}}/ n_{\text{total}}$  with $n_{\text{cur}}$ - the number of neurons in the current model, $n_{\text{total}}$ is the number of neurons in all six models, and $\mu_i$ is the average ARI score across three seeds for the current model.
We used the law of total variance and computed the variance as 
$\sigma_{\text{total}}^2 = \sum_i w_i \left[ \sigma_i^2 + \left( \mu_i - \mu_{\text{total}} \right)^2 \right]
$, where the first term captures within-dataset ARI variability and 
the second term captures between-dataset ARI variability.
\Cref{fig:perRetina} shows the ARIs per models.
We can see that the fewer neurons were present in the models the less the improvement was.

\subsection{Further analysis of monkeys data}
\label{app:monkeys}

\begin{figure}[htb]
\centering
\begin{minipage}{0.67\textwidth}
    \includegraphics[width=\linewidth]{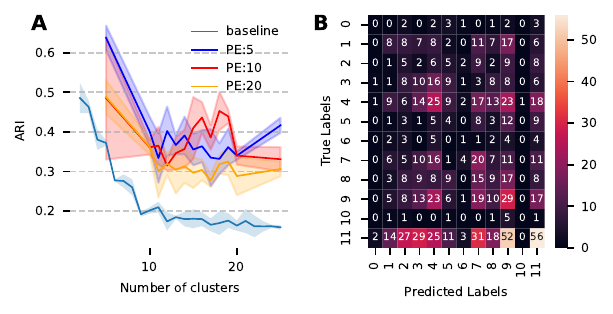}
\end{minipage}
\hspace{1em} 
\begin{minipage}{0.25\textwidth}
    \captionof{figure}{
    \textbf{A} ARI between models seeds for models trained on all 1244 neurons.
    \\\textbf{B} Confusion matrix between predictions of the model trained on 1000 neurons and labels suggested in  Willeke et al. \cite{willeke2023deep}
    }
    \label{fig:monkeysApp}
\end{minipage}
\end{figure}

For monkey V4, we trained models for 5, 10 to 20 and 25 clusters, as original work reported 12 clusters for 1000 cells.
For 144 cells there were no labels and 100 cells ahd a "not properly clustered" label. Therefore, we decided to use only the 1000 labeled cells. 
For results of models trained on all cells see \Cref{fig:monkeysApp}A.
While the trends and values are qualitatively similar to the model trained only on a 1000 neurons subset, the standard deviation corridor seems to be wider, likely due to some of the "not properly clustered" neurons being in between the distinct groups. 
Please note that the labels from Willeke et al. \cite{willeke2023deep} are rather a suggestion but not ground truth as they were not verified using independent biological measurements.
For the confusion matrix of our labels and labels from Willeke et al. \cite{willeke2023deep} see \Cref{fig:monkeysApp}A.
Same as for Burg et al. \cite{burg2024discriminativestimulifunctionalcell}, our labels do not perfectly match the ones proposed in Willeke et al. \cite{willeke2023deep}.

\subsection{Compute requirements}
\label{app:compute}
All of our models can be considered light-weight in terms of compute by modern deep learning model standards.
A single mouse retina model requires less the 10Gb of GPU memory and trains under 20 minutes of walltime.
A single mouse V1 model requires $\approx$ 12Gb of GPU memory and trains for under 2 hours of walltime.
A single marmoset retina model uses ~40Gb GPU and trains for under 16 hours of walltime.
A single monkey model requires ~24Gb of memory and trains for under 2 hours of walltime.

We use a local infrastructure cluster with 8 NVIDIA RTX A5000 GPUs with 24Gb of memory each for mouse experiments.
For mouse retina, marmoset retina, and monkey V4 we used 40Gb NVIDIA A100.

\subsection{Broader impact}
\label{app:broader}
Our work contributes to building more reproducible models, which are more suitable for making biologically meaningful statements. 
It is even more related to derive a functional taxonomy of cell types in the primary visual cortex, which can enhance our understanding of brain function and support the development of treatments for neurodegenerative diseases.

\subsection{Experimental settings}
\label{app:expSetting}
For marmoset RGC dataset, we used the three layer CNN described in \cite{Vystrcilova2024}. We trained it for a maximum of 1000 epochs, stopping early if validation correlation did not improve for 20 epochs. The learning rate of both pretraining and training with the clustering loss was initially 0.005 and reduced during training using the \texttt{ReduceLROnPlateau} learning rate scheduler, patience 10 and minimal learning rate $1e^{-8}$.
\\For SENSORIUM 2022, we used their model and training hyperparameters for the baselines training.
Pretraining duration, learning rates and clustering strength $\beta$ is reported in every experiment.
For mouse retina, we followed Hofling et al. \cite{hofling2024chromatic} model and training hyperparameters, changing only learning rate from 0.01 to 0.005 to improve baselines stability. 
Clustering strength was set to 0.001 across all experiments.
For monkey V4 data we followed model and training hyperparameters from \cite{pierzchlewicz2023energy}, again only changing the learning rate from  $3 \cdot 10^{-4}$ to $5 \cdot 10^{-5}$ to improve baselines stability. 
Clustering strength was set to 0.001 across all experiments.
Changing learning rate in both cases did not impacted performance in more than std boundaries.

\end{document}